\documentclass{aa}  

\usepackage{graphicx}
\usepackage{txfonts}
\usepackage{newtxtext,newtxmath}
\usepackage[autostyle]{csquotes}
\usepackage{subfig}
\usepackage{xcolor}
\usepackage{hyperref}
\begin{document}

   \title{Magnetic fields in the outskirts of PSZ2 G096.88+24.18 from depolarization analysis of radio relics}
   \titlerunning{Magnetic fields in the outskirts of PSZ2 G096.88+24.18 from depolarization analysis of radio relics}
   \authorrunning{E. De Rubeis et al.}

   \author{E. De Rubeis\inst{1,2},
          C. Stuardi\inst{2},
          A. Bonafede\inst{1,2},
          F. Vazza\inst{1,2,3},
          R. J. van Weeren\inst{4},
          F. de Gasperin\inst{2,3},
          M. Br{\"u}ggen\inst{3}
          }

   \institute{Dipartimento di Fisica e Astronomia, Universit\`a di Bologna, via Gobetti 93/2, I-40129 Bologna, Italy\\
              \email{emanuele.derubeis2@unibo.it}
              \and
    INAF - Istituto di Radioastronomia di Bologna, Via Gobetti 101, I-40129 Bologna, Italy
    \and
    Hamburger Sternwarte, University of Hamburg, Gojenbergsweg 112, 21029 Hamburg, Germany
    \and
    Leiden Observatory, Leiden University, PO Box 9513, NL-2300 RA Leiden, the Netherlands}

   \date{Received XXX; accepted YYY}

% \abstract{}{}{}{}{} 
% 5 {} token are mandatory
 
  \abstract
   {Radio relics are diffuse, non-thermal radio sources present in a number of merging galaxy clusters. They are characterized by elongated arc-like shapes and highly polarized emission (up to $\sim$60\%) at GHz frequencies, and are expected to trace shocks waves in the cluster outskirts induced by galaxy cluster mergers. Their polarized emission can be used to study the magnetic field properties of the host cluster.}
   {In this paper, we investigate the polarization properties of the double radio relics in PSZ2 G096.88+24.18 using the rotation measure (RM) synthesis, and try to constrain the characteristics of the magnetic field that reproduce the observed depolarization as function of resolution (beam depolarization). Our aim is to understand the nature of the low polarization fraction that characterizes the southern relic with respect to the northern relic.}
   {We present new $1-2$ GHz \textit{Karl G. Jansky} Very Large Array (VLA) observations in multiple configurations. We derive the rotation measure and polarization of the two relics by applying the RM synthesis technique, thus solving for bandwidth depolarization in the wide observing bandwidth. To study the effect of beam depolarization, we degraded the image resolution and studied the decreasing trend of polarization fraction with increasing beam size. Finally, we performed 3D magnetic field simulations using multiple models for the magnetic field power spectrum over a wide range of scales, in order to constrain the characteristics of the cluster magnetic field that can reproduce the observed beam depolarization trend.}
   {Using RM synthesis, we obtained a polarization fraction of ($18.6 \pm 0.3$)\% for the norther relic and ($14.6 \pm 0.1$)\% for the southern one. Having corrected for bandwidth depolarization, and after noticing the absence of relevant complex Faraday spectrum, we infer that the nature of the depolarization for the southern relic is external, and possibly related to the turbulent gas distribution within the cluster, or to the complex spatial structure of the relic. The best-fit magnetic field power spectrum, that reproduces the observed depolarization trend for the southern relic, is obtained for a turbulent magnetic field model, described by a power spectrum derived from cosmological simulations, and defined within the scales of $\Lambda_{\rm{min}}=35~\rm{kpc}$ and $\Lambda_{\rm{max}}=400~\rm{kpc}$. This yields an average magnetic field of the cluster within 1$~\rm{Mpc}^3$ volume of $\sim 2~\rm{\mu G}$.}
   {}


   \maketitle
%
%-------------------------------------------------------------------

\section{Introduction}
\label{sec:intro}

\par Mergers of galaxy clusters are among the most energetic events in the Universe, in which a fraction of the  kinetic energy is dissipated into the acceleration of charged relativistic particles, within the hot thermal plasma that fills the clusters, the intracluster medium (ICM). Merger shocks are usually launched along the merger axis, with Mach numbers in the range $M \sim 2-5$~\citep[e.g.,][]{2010vanweeren,akamatsu2013,2017wittor,botteon2020}. Despite their low $M$, these shocks seem to be able to (re)accelerate particles that, with the $\mu$G magnetic fields that characterize the clusters environment~\citep[e.g.,][]{2018vacca}, emit synchrotron radiation at radio frequencies. These structures, known as \enquote{radio relics}, represent extraordinary examples of the consequences of the propagation of shocks through the cluster after a cluster-cluster merger~\citep[for reviews see][]{feretti2012,2019vanweeren}.
\par Radio relics are mostly found in the outskirts of galaxy clusters and are characterized by elongated shapes, with lengths of 0.5-2 Mpc, and strong polarization at GHz frequencies~\citep[$\gtrsim 20\%$, e.g.,][]{1998enblin,2022stuardi}. Relics are among the most polarized sources in the extragalactic sky: the polarization fractions can reach $\sim 60\%$ in some cases. This high degree of polarization is expected in sources that trace edge-on shock waves~\citep{1998enblin}. These source properties suggest the presence of large-scale magnetic fields that have intensities of $0.1-10~\rm{\mu G}$, tangled on scales ranging from few to hundreds of kpc~\citep{2013bruggen}.
\par Another way in which the presence of the cluster magnetic field is unveiled is the Faraday rotation effect which affects the propagation of linearly polarized radiation through the magnetized ICM. This effect causes the rotation of the polarization angle, $\chi$, by a quantity that depends on the wavelength squared of the emission, $\lambda^{2}$, and on the Faraday depth
\begin{equation}
\label{eq:faradaydepth}
    \phi = 0.812 \int_{\rm{source}}^{\rm{observer}} n_{e}B_{\parallel} \, dl \quad \rm{rad~m^{-2}}~,
\end{equation}
where $n_{e}$ is the thermal electron density in $\rm{cm^{-3}}$, $B_{\parallel}$ is the magnetic field component along the line of sight in $\rm{\mu G}$, and $dl$ is the infinitesimal path length in pc~\citep{1966burn}. When the relation between $\lambda^2$ and $\chi$ is linear, $\phi$ is traditionally called rotation measure (RM).
\par The Faraday rotation effect causes a loss of observed polarized intensity with respect to the intrinsic one, depending on $\phi$ and the observing wavelength, because of the mixing of different polarization angles along the line of sight. There are several depolarization mechanisms that can arise from physical and/or instrumental effects. Throughout the paper we focus particularly on the beam depolarization, caused by the beam-size, and the bandwidth depolarization, which is an instrumental effect due to the large observational bandwidth; a more comprehensive description of these mechanisms can be found in Section~\ref{sec:depolarization_mechanisms}.
\par Information about the magnetic field in individual clusters through RM studies has been obtained for about 30 objects, including both merging and relaxed clusters. These studies have led to a great improvement in our knowledge of clusters magnetic fields~\citep[e.g.,][]{2010bonafede, 2021stuardi, 2021digennaro, 2022rajpurohit, 2022degasperin}. The magnetic field power spectrum is not well known, and can be assumed to follow a Kolmogorov power spectrum ($E_{B} \propto k^{-11/3}$ in 3D) from the velocity distribution. However, a power-spectrum distribution of magnetic field fluctuations is not expected in any dynamo theory~\citep[e.g.,][and references therein]{2019JPlPh..85d2001R} and recent magneto-hydrodynamical (MHD) simulations~\citep{2018vazza,2019dominguezfernandez} of galaxy clusters have shown that the magnetic power spectrum is more complex than a simple power-law spectrum because of the dynamics of the ICM. Hence, RM studies represent an important way to determine the characteristics of the ICM magnetic field, in both intensity and structure along the line-of-sight.
\par In this paper, we studied the polarization properties of a double radio relics system found in the merging cluster PSZ2 G096.88+24.18~\citep{2011pszcatalogue}. The host cluster has been found to be originated by the merger between two equal mass sub-clusters with merger axis along the sky plane~\citep{2021finner, tumer2023}. In double radio relics systems we expect high levels of fractional polarization due to the plane-of-the-sky, randomly oriented magnetic field compression operated by the passing shock wave. This shock wave is expected to have similar properties in both its propagation directions, so we also expect the two originating relics to show similar polarization properties~\citep[see Tab. 4 in][]{2022stuardi}. In this cluster, however, the southern relic shows a lower polarization fraction with respect to the northern one and to what is typically found in radio relics, as reported by previous work~\citep[i.e.][]{2014degasperin,2021jones}, with values below $10\%$. Previous polarization studies have been performed integrating over the whole band so that bandwidth depolarization effect must be present (see Sec.~\ref{sec:depolarization_mechanisms}). With our analysis we studied the polarization using the rotation measure synthesis technique: minimizing the effect of bandwidth depolarization, we aim to understand the nature of the low polarization fraction that characterizes the relic south of the cluster.
We used new and archival \textit{Karl G. Jansky} Very Large Array (VLA) observation in the L-band (1-2 GHz) in three different configurations, in order to achieve both high angular resolution and high sensitivity to extended emission. We use the rotation measure synthesis technique and by doing this we are able to recover higher fractional polarization with respect to the integrated polarization analysis, especially in some regions of the southern relic. We evaluate the beam depolarization within the relics degrading the resolution of our images, obtaining a depolarization trend for the selected regions. We then use 3D MHD simulations to constrain the magnetic field characteristics, like turbulence scales or mean magnetic field, that can reproduce the observed depolarization of the southern relic.
\par This paper is organized as follows: In Section~\ref{subsec:psz2g096} we present the main peculiarities of PSZ2 G096.88+24.18. In Section~\ref{sec:data_analysis} we detail the VLA observations and their data products. In Section~\ref{sec:pol_analysis} the polarization analysis is described with our results. The simulations results are then presented in Section~\ref{sec:simulations}. The discussion and conclusions are in Section~\ref{sec:discussion}. In this paper we assume a flat $\rm{\Lambda}CDM$ cosmology, with $ H_{0} = 70~\rm{km~s^{-1}~Mpc^{-1}}$ and $\Omega_{M} = 0.3$. At the redshift of PSZ2 G096.88+24.18 ($z$ = 0.304), 1" corresponds to a linear scale of 4.45 kpc.
\begin{figure*}[ht!]
\centering
 \subfloat{\includegraphics[width=\columnwidth]{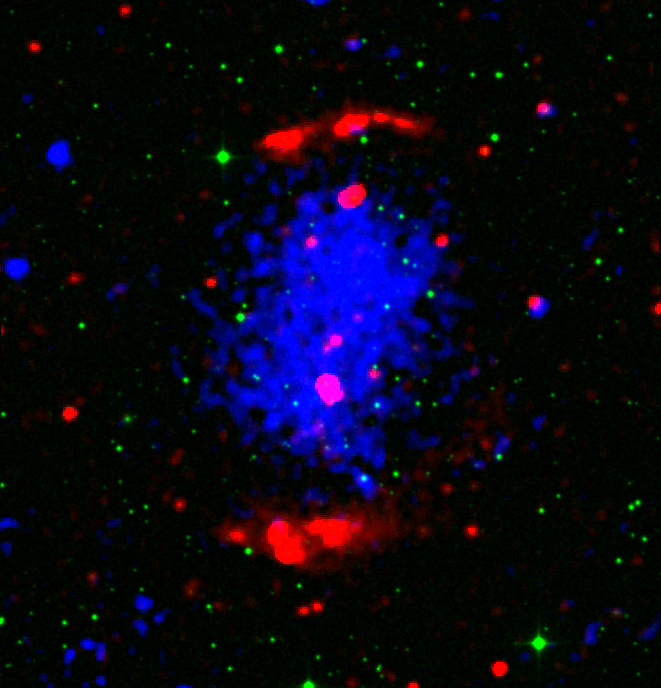}}
 \subfloat{\includegraphics[width=\columnwidth]{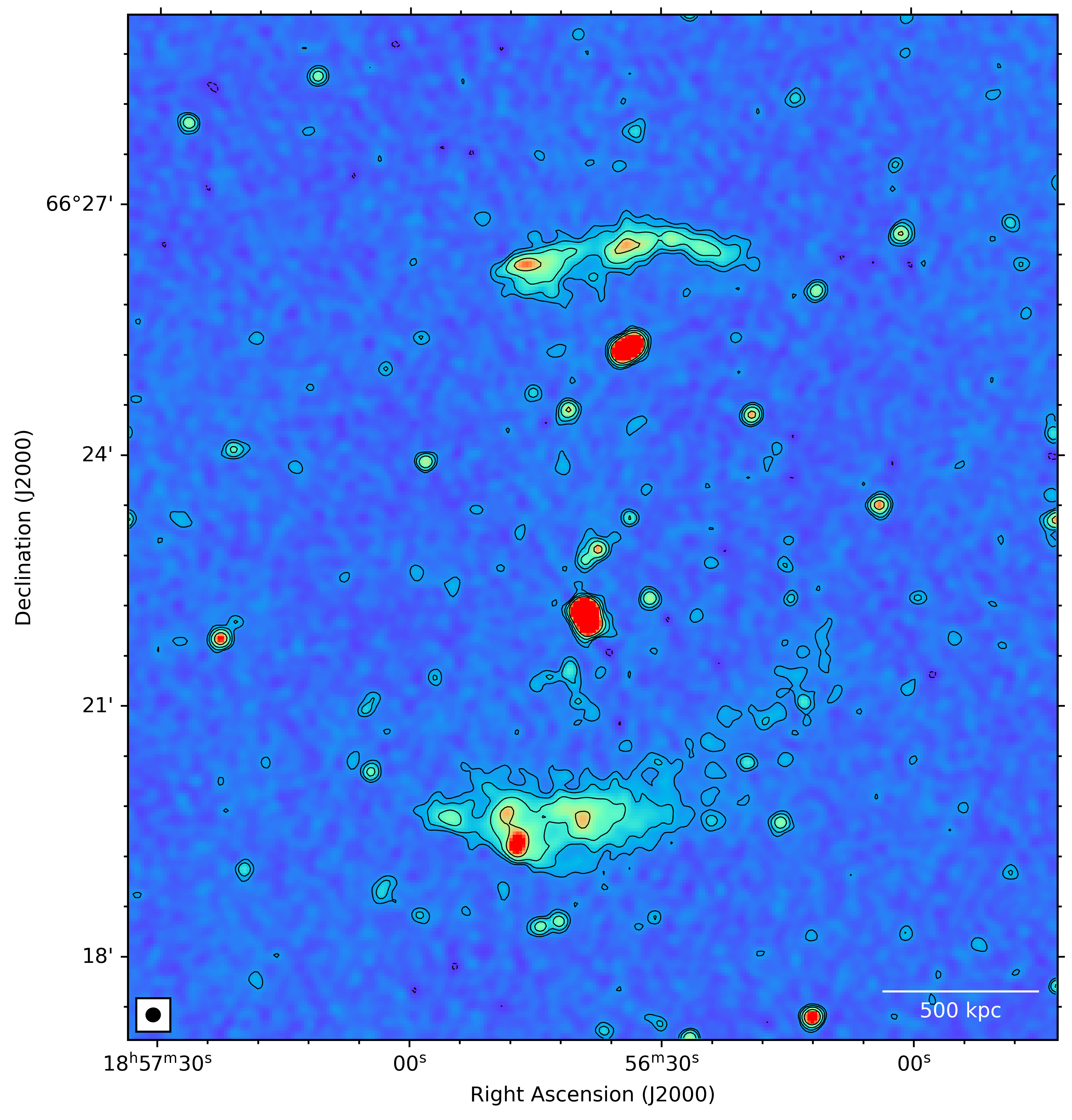}}
 \caption{\textit{Left}: composite optical (DSS2, green) + radio total intensity (Stokes $I$, red) + XMM-Newton (blue) image of PSZ2 G096.88+24.18. \textit{Right}: radio total intensity image obtained with VLA in the L-band (resolution $9.6" \times 9.2"$, RMS noise $\sigma_{\rm{rms}}=12~\rm{\mu Jy~beam^{-1}}$) combining B, C, and D configurations. The contour levels are drawn at $-3\sigma$, $3\sigma$, and then increase by a factor of 2. The beam is shown with a black circle in the bottom left corner.}
 \label{fig:PSZ096_totalintensity_10asec}
\end{figure*}

\section{PSZ2 G096.88+24.18}
\label{subsec:psz2g096}
PSZ2 G096.88+24.18~\citep[Fig.~\ref{fig:PSZ096_totalintensity_10asec}, also known as ZwCL1856.8+6616,][]{1961zwicky, 2011pszcatalogue} was detected in the Planck Sunyaev-Zel'dovich Survey~\citep{2011pszcatalogue}, and reported to have a redshift $z=0.304$ and mass $M$\footnote{\label{note1} $M_{500(200)}$ is the mass enclosed within the radius $r_{500(200)}$, within which the mean density of the cluster is $500(200)$ times the critical density of the Universe at the cluster redshift.}$_{500}=(4.7 \pm 0.3)\times 10^{14}~M_{\odot}$ in the second Planck data release~\citep{2016psz2catalogue}. \\
Using optical and X-ray data, \citet{2021finner} investigated the merger scenario of PSZ2 G096.88$+$24.18. They estimated a 1:1 mass ratio, with a total mass of $M_{200} = 2.4_{-0.7}^{+0.9} \times 10^{14}~M_{\odot}$, from simultaneously fitting two Navarro-Frenk-White (NFW) halos to the lensing signal. Combined with the spectroscopic results of~\citet{2019golovich}, who found a single-peak redshift distribution of galaxies associated with PSZ2 G096.88+24.18, the merger is likely a head-on collision on the plane of the sky and the time since collision is $0.7^{+0.3}_{-0.1}~\rm{Gyr}$. The dynamical status of the system is supported by X-ray observations, as the cluster shows two clumps of gas at a projected distance of $\sim 600~\rm{kpc}$.
\par \citet{2014degasperin} discovered a pair of radio relics on the northern and southern edges of PSZ2 G096.88+24.18 at 1.4 GHz with Westerbork Synthesis Radio Telescope (WSRT). The two relics have dimensions of $\sim$900 kpc (northern) and $\sim$1.4 Mpc (southern), and are at a distance from the midpoint between the two X-ray peaks of emission of 770 kpc and 1145 kpc, respectively. From a polarization analysis, they found that the relic N has electric field vectors mostly aligned perpendicular to the relic extension meaning that the magnetic field is aligned with the relic, with a fractional polarization up to $\sim30\%$. However, the relic S shows vectors that are $\sim$45° apart from being perpendicular to the relic extension, with a lower polarization fraction that reaches values of about $10\%$.
\par \citet{2021jones} studied this cluster using LOw Frequency ARray~\citep[LOFAR,][]{vanhaarlem2013} High Band Antenna (HBA) observations, at 120-187 MHz, and VLA L-band radio observations (with C and CnB-B configurations), as well as Chandra X-ray observations of the cluster. The morphology, location at the cluster periphery, and spectral index variation of the arc-like radio structures in PSZ2 G096.88+24.18 confirm that we are observing a pair of radio relics. At 144 MHz the relics have largest linear sizes (LLS) of $\sim$0.9 and 1.5 Mpc and a flux ratio of 1:3.5 for the north and south relics respectively. Both relics have a non-uniform brightness along their major axis, and high resolution images show filament-like substructures in the northern relic. Both the north and south relics exhibit spectral steepening from the shock edge towards the cluster center, that is expected from downstream synchrotron and inverse Compton (IC) losses, as observed in many other relics~\citep[e.g.,][]{2012bonafede,2016vanweeren,2017hoang}. In their polarimetric analysis, they found that in the northern relic there are significant regions of polarized emission, mostly corresponding to the brightest parts of the relic, with a linear polarization fraction that ranges from 10\% to 60\% and magnetic field ordered and compressed along the shock front, as a consequence of the shock passage~\citep{1998enblin}. The southern relic shows polarized emission in only a few small regions and the polarization fraction is much lower than in the northern relic, reaching a maximum of 20\%. In these small regions, the electric field vectors do not lie perpendicular to the shock, in agreement with the results by~\citet{2014degasperin} although obtained with a different dataset. Despite the availability of X-rays data, they were unable to detect evidence of a shock front at the position of the relics. Due to the low count statistics in cluster outskirts, the Chandra data are likely not sensitive enough to detect a shock.~\citet{2021finner} also did not detect a shock with a 12 ks XMM-Newton observation. Recently,~\citet{tumer2023} found an indication of a shock front coinciding with the northern relic with Chandra and NuSTAR spectra, but deeper exposures are required to draw more accurate surface brightness profiles that can confirm it. 

\begin{table}
\centering
\caption{Properties of PSZ2 G096.88+24.18 from literature.}
\label{tab:cluster_properties}
\begin{tabular}{lr}
\hline\hline
Property & Value \\
\hline
$z$ & $0.304$ \\
R.A. (h m s) & 18, 56, 38 \\
Dec. ($\degr$ ' ") & 66 23 10 \\
$M_{500}~(10^{14}\rm{M_{\odot}})$ & $4.7 \pm 0.3$ \\
%$L_{500}~(10^{44}~\rm{erg~s^{-1}})$ & $(11.5 \pm 1.9)\%$ \\
$kT~(\rm{keV})$ & $3.7_{-0.5}^{+0.6}$ \\
%$P_{1.4}~(10^{23}~\rm{W~Hz^{-1}})$ & $(11.0 \pm 6.1)\%$ \\
\hline
\end{tabular}
\end{table}

%%%%%%%%%%%%%%%%%%%%%%%%%%%%%%%%%%%%%%%%%%%%%%%%%%%%%%%%%%%%%%%%%%%%%%%%%%%

% IMAGING

%%%%%%%%%%%%%%%%%%%%%%%%%%%%%%%%%%%%%%%%%%%%%%%%%%%%%%%%%%%%%%%%%%%%%%%%%%%

\section{Data analysis}
\label{sec:data_analysis}

\subsection{Calibration}

We used both new and archival observations of PSZ2 G096.88+24.18 performed with VLA in the L-band (1-2 GHz). We analyzed a total of 7.5 hours of observations: 4 hours in B-configuration, 3 hours in C-configuration and 0.5 hours in D-configuration (see also Tab.~\ref{tab:obs}). With respect to previous studies, which only used C- and CnB-configurations, our observations are sensitive to a broader range of angular scales.
\par The dataset were pre-processed with the VLA \texttt{CASA}\footnote{\url{https://casa.nrao.edu/}}~\citep{2007casa} calibration pipeline. This pipeline is optimized for Stokes I continuum data and it performs standard flagging and calibration procedures. We used the \texttt{CASA 5.6.1} package to complete the calibration also for the cross-correlation polarization products and to perform additional flagging. In particular, we used the \texttt{rflag} algorithm, gradually decreasing the flagging threshold from 5$\sigma$ to 2$\sigma$ in both time and frequency when necessary. We also applied manual flagging to flag entire spectral windows. We derived final delay, bandpass, gain/phase, leakage and polarization angle calibrations. We used the \citet{Perley13} flux density scale for wide-band observations as a model for the primary calibrator of each observation (3C286). We used as phase calibrators: J1849+6705 (B-Configuration), J2022+6136 (C-configuration), and J1748+7005 (D-configuration). An unpolarized source was used to calibrate the on-axis instrumental leakage (J1407+2827 for B- and D-configurations, and QSO B1404+286 for the C-configuration). The primary calibrator was also used for the absolute polarization angle calibration. We made a polynomial fit to its known values of linear polarization fraction and polarization angle to build a frequency-dependent polarization model, following the NRAO polarimetry guide\footnote{\url{https://science.nrao.edu/facilities/vla/docs/manuals/obsguide/modes/pol}}. The final calibration tables were applied to the target.
\par Radio frequency interference (RFI) was removed manually and using statistical flagging algorithms also from the cross-correlation products (RL, LR). The calibrated data were then averaged in time down to 10 s and in frequency to the channel width of 4 MHz, in order to speed up the subsequent imaging and self-calibration processes.

\subsection{Total intensity imaging}

For the total intensity imaging and the self-calibration we used the \texttt{CASA v6.4.0.16} package. Data have been imaged using the \texttt{tclean} task of \texttt{CASA} with the multi-scale multi-frequency imaging deconvolution algorithms, in order to better reconstruct the extended structure of the source. Moreover, we used three orders in Taylor expansion~\citep[setting \texttt{nterms}=3,][]{2011rau} to take into account, both, the source spectral index and the primary beam response at large distances from the pointing center. Given the wide-field to be imaged ($\sim 28' \times 28'$), we used also the $w$-projection algorithm~\citep{2008cornwell} to correct for the sky curvature and better clean the most peripheral sources of the image, reducing the artifacts: we imposed \texttt{wprojplanes}=256. For all the configurations we used a Briggs weighting scheme~\citep{1995briggs}, with a robustness parameter of 0. For the sole D-configuration we performed 4 self-calibration cycles, in order to improve the image quality.
\par Once the different configurations have been inspected individually, we have combined the observations in the visibility space, in order to obtain a single dataset over which we will perform our polarimetric analysis. From this final combined dataset, we subtracted the sources outside a region of about 12' radius in order to focus only on the cluster region and speed up the computational time in the following analysis. We note that, by restricting the subsequent polarization analysis to the central region of the field, we ensure that the polarization leakage due to instrumental effects is below the 2$\%$ level. The final image has a restoring beam of $9.6" \times 9.2"$, using a Gaussian taper of $10"$, and noise $12~\rm{\mu Jy~beam^{-1}}$ (Fig.~\ref{fig:PSZ096_totalintensity_10asec}).

\begin{table*}[h!]
\centering
\caption{List of the observations used in this paper.}
\begin{tabular*}{\textwidth}{c @{\extracolsep{\fill}} lcccccr}
\hline\hline
Freq. & P.I. & Array Conf. & Obs. date & Tot. obs. time & Beam Size & RMS \\
(GHz) & & & & & & ($\rm{Jy~beam^{-1}}$)\\
\hline
$1.5$ GHz & Bonafede & B & 2017 Oct. \& Dec. & 4h 20m & $3.6" \times 2.7"$ & $1.4 \times 10^{-5}$ \\
\hline
$1.5$ GHz & de Gasperin & C & 2016 Feb. & 3h 30m & $14.5" \times 9.7"$ & $1.9 \times 10^{-5}$ \\
\hline
$1.5$ GHz & Bonafede & D & 2017 Feb. & 46m & $61.3" \times 29.3"$ & $6.8 \times 10^{-5}$ \\
\hline
\end{tabular*} 
\tablefoot{Column 1: central frequency of the observation. Column 2: principal investigators. Column 3: VLA antennas configuration. Column 4: year and months of observation. Column 5: size of the restoring beam with Briggs weighting scheme and robust=0. Column 6: best RMS noise obtained for each dataset.}
\label{tab:obs}
\end{table*}

\section{Polarization analysis}
\label{sec:pol_analysis}

\subsection{Polarization imaging}
\label{subsec:polarizationimaging}
Once our combined dataset has been obtained, we have to prepare the way for the RM synthesis technique. We used \texttt{WSClean v3.1}~\citep[$w$-stacking clean,][]{2014offringa} to produce images in Stokes $I$, $Q$, and $U$. With this software, it is possible to make both a multi-frequency synthesis (MFS) image and a cube with an image made for each channel, required for the RM synthesis using the \texttt{CIRADA RM-tools v1.2.0}\footnote{\label{note2} \url{https://github.com/CIRADA-Tools/RM-Tools}}~\citep{2020rmtools}. We produced a cube made by 64 images for $I$, $Q$, $U$ Stokes, as well as a wide-band image. 
\par We used 64 images for each cube with a bandwidth of 16 MHz per image. This choice is motivated by the fact that, within each channel, we are sensitive to a maximum observable $\phi_{\rm{max}} \sim 655~\rm{rad~m^{-2}}$ (see Sec.~\ref{subsec:mfsrmsynthesisresults}): this means that we do not have significant bandwidth depolarization within the channels, since RM in non-cool-core clusters are generally below $500~\rm{rad~m^{-2}}$~\citep{2016bohringer}. The three Stokes images have been cleaned separately, using \texttt{join-channels} for the multi-frequency deconvolution. This performs peak finding in the squared sum over all output channels, but subtracts components independently from the channels. We used also the Briggs weighting scheme with robust=0. For the imaging cube, we forced all the 64 images to have the same restoring beam, corresponding to the one of the first sub-band (the lowest resolution one, that is $13"$). We did this also for the MFS images, in order to be consistent for the comparison of the obtained results. Then, the MFS images as well as the $I$, $Q$, and $U$ Stokes imaging cubes, are corrected for the primary beam (PB). In \texttt{CASA}, we can use the task \texttt{widebandpbcor} in order to compute a set of PBs at the specified frequencies, in our case 64 PBs, corresponding to the ones that constitute the imaging cubes. Lastly, within the 64 images of the cubes, 8 were removed because the frequencies between 1.51 and 1.65 GHz are completely flagged because of the strong RFI. The final PB-corrected cubes are then composed of 56 images.

\subsection{Integrated polarization results}
\label{subsec:mfsintegratedresults}

Before performing the RM-synthesis we decided to evaluate the polarization fraction of the relics integrated over the full band. This will serve as a reference value to evaluate the depolarization due to bandwidth-averaging. To do this, we used the MFS images produced with \texttt{WSClean} and corrected for the primary beam. We evaluated the integrated flux in P ($S_{\rm{P,int}}$) and in total intensity $I$ ($S_{\rm{I,int}}$). We selected two regions, one for each relic, within which we calculated the polarization fraction. These regions have been selected following the $5\sigma$ contour for both relics in the total intensity image, where $\sigma$ is the RMS noise of the total intensity image, and are shown in Fig.~\ref{fig:PSZ096_relics_regions}.
\par The polarization image has been obtained combining the $Q$ and $U$ images and then correcting for the Ricean bias following~\citet{2012george}, in which the intrinsic polarized intensity for each pixel is
\begin{equation}
\label{eq:riceanbias}
P =\sqrt{{P_{\rm{obs}}}^{2}-2.3\sigma_{P_{\rm{obs}}}^{2}}.
\end{equation}
$P_{\rm{obs}}=\sqrt{Q_{\rm{obs}}^{2}+U_{\rm{obs}}^{2}}$ is the observed polarized flux, while the value $\sigma_{P_{\rm{obs}}}$ has been computed considering the average RMS noise level between the $Q_{\rm{obs}}$ and $U_{\rm{obs}}$ images and is $3 \times 10^{-6}~\rm{Jy~beam^{-1}}$. For each region, the flux density in both total intensity $I$ ($S_{\rm{I,int}}$) and polarized intensity $P$ ($S_{\rm{P,int}}$) has been obtained, as well as the respective statistical flux error calculated as
\begin{equation}
\label{eq:fluxerror}
\delta S_{\rm{flux,int}} = \frac{\sigma_{\rm{flux,rms}}}{\sqrt{n_{\rm{beam}}}},
\end{equation}
where flux = $I$, $P$, and $n_{\rm{beam}}$ is the number of beams within the chosen region. Here, we ignore the systematic error on the flux scale, as we are interested in the relative polarization fraction. The measured flux densities and their uncertainties are listed in Tab.~\ref{tab:fluxes}.
\par We calculate the integrated polarization fraction for both regions as 
\begin{equation}
\label{eq:polarizationfraction}
f_{\rm{pol,int}} = \frac{S_{\rm{P,int}}}{S_{\rm{I,int}}},
\end{equation}
and the error on the polarization fraction is obtained through the propagation of errors
\begin{equation}
\label{eq:polfracerr}
\delta f_{\rm{pol,int}} = \sqrt{\Biggl(\frac{1}{S_{\rm{I,int}}} \delta S_{\rm{P,int}} \Biggr)^{2} + \Biggl(-\frac{S_{\rm{P,int}}}{S_{\rm{I,int}}^{2}} \delta S_{\rm{I,int}} \Biggr)^{2}}.
\end{equation}
For the selected regions, given the flux densities listed in Tab.~\ref{tab:fluxes} for $I$ and $P$, respectively, we obtain the fractional polarization from the integrated analysis $f_{\rm{pol,int}}$ of
\begin{itemize}
\item $(12.1 \pm 0.2)\%$ for the north region
\item $(5.8 \pm 0.1)\%$ for the south region.
\end{itemize}
These values are compatible with the results obtained by of~\citet{2021jones} who found very low levels of fractional polarization for the southern relic ($\leq 10\%$).

\subsection{Depolarization mechanisms}
\label{sec:depolarization_mechanisms}

In the presence of a magneto-ionized screen with a homogeneous distribution of thermal electrons and magnetic field, the complex polarized intensity of synchrotron radiation affected by Faraday rotation is
\begin{equation}
    \label{eq:pol_intensity_depolarization}
    P_{\rm{obs}}(\lambda) = P_{\rm{int}}e^{2i(\chi_{0}+\phi \lambda^2)}~,
\end{equation}
where $P_{\rm{int}}$ is the intrinsic polarization of the synchrotron emission, $\chi_{0}$ is the relative intrinsic polarization angle at the source of emission and $\phi$ quantifies the Faraday rotation caused by the foreground magneto-ionic medium. However, the observed polarization intensity, $P_{\rm{obs}}(\lambda)$, can be significantly lower with respect to the intrinsic value, $P_{\rm{int}}$ in the presence of more complex Faraday screens. The polarized flux can be reduced by the mixing of different polarization angles along the line of sight. Several instrumental effects can also depolarize the observed radiation. Depolarization towards longer wavelengths can occur due to the mixing of the emitting and rotating media, as well as from the finite spatial resolution of our observations~\citep{2012osullivan}. Here we present the main depolarization mechanisms, following the discussions in~\citet{1998sokoloff} and~\citet{2022stuardi}:
\begin{enumerate}
\item \textit{Differential Faraday rotation}: This effect occurs when the emitting and rotating regions are co-spatial and in a regular magnetic field. The polarization plane of the emission at the far side of the region undergoes a different amount of Faraday rotation compared to the polarized emission coming from the near side, causing depolarization when summed over the entire region. For a uniform slab, we have
\begin{equation}
P_{\rm{obs}} = P_{\rm{int}} \frac{\sin{\phi \lambda^{2}}}{\phi \lambda^{2}} e^{2i(\chi_{0}+\frac{1}{2}\phi \lambda^{2})}.
\end{equation}
We note that depolarization increases with wavelength.
\item \textit{Internal Faraday dispersion}: This occurs when the emitting and rotating regions are co-spatial and contain a turbulent magnetic field. In this case, depolarization occurs because the plane of polarization experiences a random walk through the region. For identical distributions of all the constituents of the magneto-ionic medium along the line of sight, it can be described by 
\begin{equation}
P_{\rm{obs}} = P_{\rm{int}} e^{2i\chi_{0}} \Biggl(\frac{1-e^{2i\phi \lambda^{2}-2\zeta_{\rm{RM}}^{2} \lambda^{4}}}{2\zeta_{\rm{RM}}^{2} \lambda^{4}-2i \phi \lambda^{2}}\Biggr)~,
\end{equation}
where $\zeta_{RM}$ is the internal Faraday dispersion of the medium.
\item \textit{External Faraday dispersion/beam depolarization}: This occurs in a purely external non-emitting Faraday screen. In the case of turbulent magnetic fields, depolarization occurs when many turbulent cells fall within the synthesized telescope beam. On the other hand, for a regular magnetic field, any variation in the strength or direction of the field within the observing beam will lead to depolarization. Both effects can be described by:
\begin{equation}
\label{eq:observedpolarization}
P_{\rm{obs}} = P_{\rm{int}}e^{-2\sigma_{\rm{RM}}^{2} \lambda^{4}} e^{2i(\chi_{0}+\rm{RM} \lambda^{2})},
\end{equation}
where $\sigma_{\rm{RM}}$ is the dispersion about the mean RM across the source on the sky. In this case, the depolarization increases with increasing beam-size.
\item \textit{Bandwidth depolarization}: It occurs when a significant rotation of the polarization angle of the radiation is produced across the observing bandwidth and the polarization fraction is computed averaging over the band.
\end{enumerate}

\subsection{Rotation measure synthesis}
\label{subsec:RMsynthesis}

In this Section we present the basic concepts about the RM synthesis technique, which we used to evaluate the polarized flux of the two radio relics. RM synthesis allows to recover the value of the Faraday depth, $\phi$ (see Eq.~\ref{eq:faradaydepth}), and study the polarized emission from a source. We refer to~\citet{2005brentjens} for a detailed  description of this procedure. The simplest example of Faraday rotation is given by a single source along the line of sight in a uniform rotating medium: in this case, the observed polarization vector can be written as Eq.~\ref{eq:pol_intensity_depolarization}, and the corresponding observed polarization angle is given by
\begin{equation}
\chi(\lambda^{2})=\chi_{0}+\lambda^{2}\phi.
\end{equation}
The Faraday depth $\phi$ coincides with the RM when there is no internal Faraday rotation and there is only one source along the line of sight~\citep{2005brentjens}. As we will see in Sec.~\ref{subsec:mfsrmsynthesisresults}, this is also the case for the relics studied in this paper.
\par
\begin{figure}
\subfloat{\includegraphics[scale=0.55]{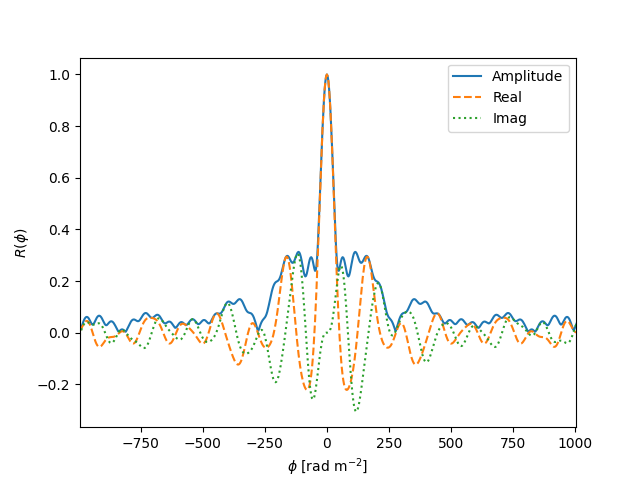}} \\
\subfloat{\includegraphics[scale=0.55]{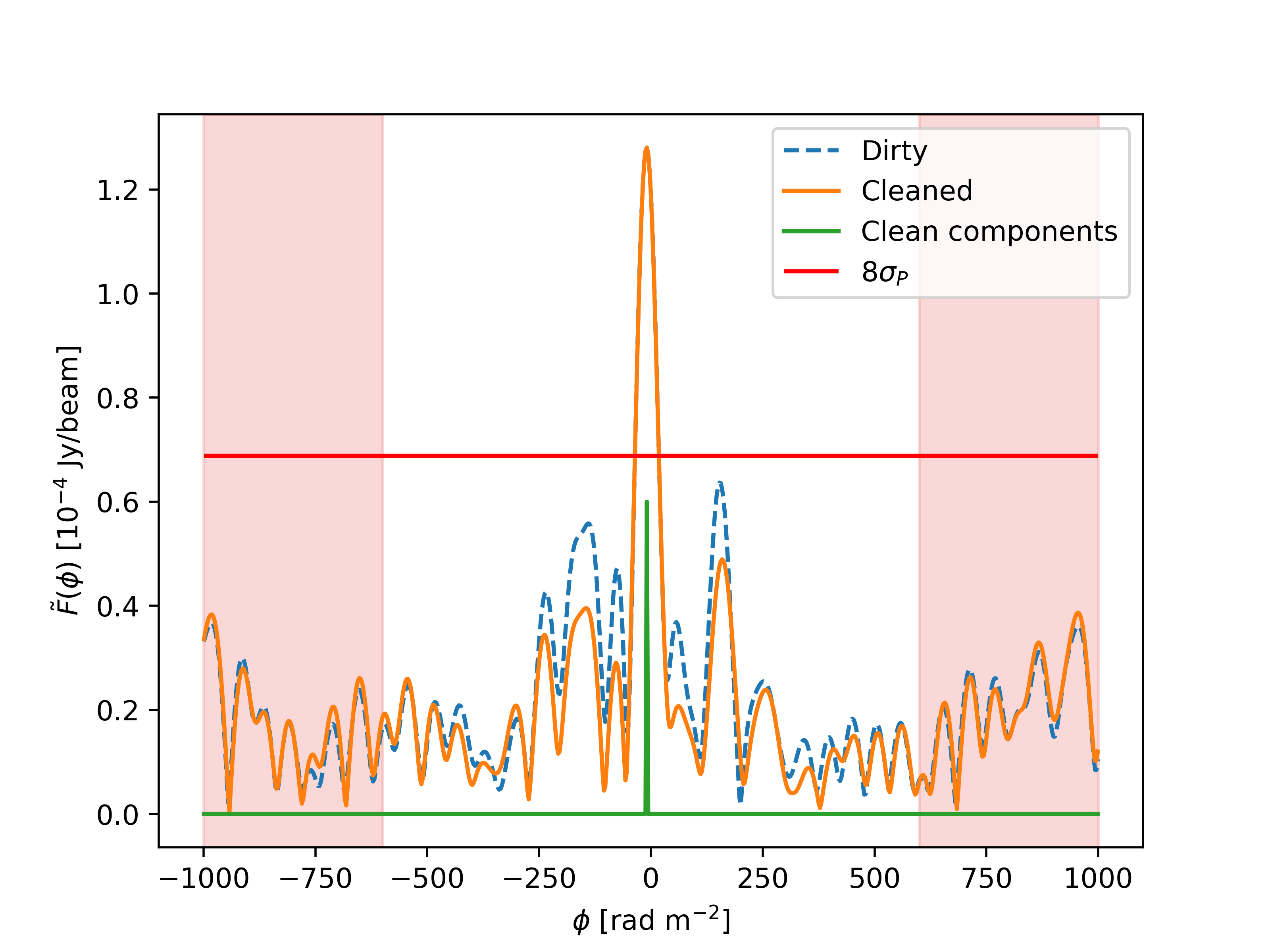}}
\caption{\textit{Top}: RMSF for our observations. The dashed orange and dotted green represent the real and imaginary components, respectively, while the blue line shows the total amplitude.\\
\textit{Bottom}: Faraday spectrum for one pixel of the northern relic. We show here both the dirty (dashed blue) and the cleaned (orange line) spectra. The horizontal red line that represents the threshold used for RM clean, and in green we have the clean component selected by the cleaning algorithm. In light red we show the range of $\phi$ used to evaluate $\sigma_{S_{\rm{P_{rm,obs}}}}^{2}$.}
\label{fig:rmsf_and_fdf}
\end{figure}

 The rotation measure synthesis allows us to solve for the so-called \enquote{n$\pi$ ambiguity}, correct for bandwidth depolarization and to recover low signal-to-noise sources. Furthermore,~\citet{1966burn} found a Fourier transform relationship between the observed polarized flux and the polarized flux expressed as a function of the Faraday depth which is still valid when the relationship between $\chi$ and $\lambda^2$ is non-linear.~\citet{2005brentjens} extended the work of~\citet{1966burn} introducing a weight function (also known as window function), $W(\lambda^{2})$, which is non-zero only at values of $\lambda^{2}$ which are sampled by the telescope. With this implementation, we can rewrite the observed polarized flux density as
\begin{equation}
\label{eq:obspol}
\tilde{P}(\lambda^{2}) = W(\lambda^{2})P(\lambda^{2}) = W(\lambda^{2}) \int_{-\infty}^{+\infty} F(\phi) e^{2i \phi(\lambda^{2}-\lambda_{0}^{2})} \, d\phi
\end{equation}
and the \enquote{reconstructed} Faraday dispersion function, also known as \enquote{Faraday spectrum} (Fig.~\ref{fig:rmsf_and_fdf}), as
\begin{equation}
\label{eq:reconstructedfdf}
\tilde{F}(\phi) = F(\phi) \circledast R(\phi) = K \int_{-\infty}^{+\infty} \tilde{P}(\lambda^{2})e^{-2i \phi (\lambda^{2}-\lambda_{0}^{2})} \, d\lambda^{2}~,
\end{equation}
where $K$ is the inverse of the integral over $W(\lambda^{2})$, the $\circledast$ denotes convolution, and $R(\phi)$ is the so-called RM transfer function (RMTF) or RM spread function (RMSF, Fig.~\ref{fig:rmsf_and_fdf}) .
\begin{equation}
\label{eq:rmsf}
R(\phi) \equiv K \int_{-\infty}^{+\infty} W(\lambda^{2}) e^{-2i \phi (\lambda^{2}-\lambda_{0}^{2})} \, d\lambda^{2}.
\end{equation}
In this  work, the latter denomination will be used.
\par The quantity $\lambda_{0}^{2}$ is the mean of the sampled $\lambda^{2}$ values, weighted by $W(\lambda^{2})$. $\tilde{F}(\phi)$ is $F(\phi)$ convolved with $R(\phi)$, therefore after Fourier filtering by weight function $W(\lambda^{2})$. The quality of reconstruction depends mainly on the weight function $W(\lambda^{2})$. Hence, a complete and wide range of $\lambda^{2}$ measurements leads to a better defined RMSF and a fine sample in $\phi$ space, allowing a better reconstruction of $F(\phi)$.~\citet{2005brentjens} showed that Eqs.~\ref{eq:reconstructedfdf} and~\ref{eq:rmsf} can be approximated as sums.
%\begin{align}
%\tilde{F}(\phi) & \approx K \sum_{c=1}^{N} \tilde{P}_{c} \, e^{-2i \phi (\lambda_{c}^{2} - \lambda_{0}^{2})}, \\
%R(\phi) & \approx K \sum_{c=1}^{N} W_{c} \, e^{-2i \phi (\lambda_{c}^{2} - \lambda_{0}^{2})}, \\
%K &= \Biggl(\sum_{c=1}^{N} W_{c} \Biggr)^{-1},
%\end{align}
%where the index $c$ refers to the individual frequency channels in which the polarized flux is observed at the radio telescope and $N$ is the total channels number.
\par Three main parameters are involved when using the RM synthesis technique: the channel width $\delta \lambda^{2}$, the width of the $\lambda^{2}$ distribution $\Delta \lambda^{2}$, and the shortest wavelength squared $\lambda_{\rm{min}}^{2}$. These parameters determine, respectively, the maximum observable Faraday depth, the resolution in $\phi$ space, and the largest scale in $\phi$ space to which one is sensitive. Given the parameters of our L-band observations, the FWHM of the main peak of the RMSF in our data is given by
\begin{equation}
\label{eq:deltaphi}
\delta \phi \approx \frac{2\sqrt{3}}{\Delta \lambda^{2}} \sim 53~\rm{rad~m^{-2}}
\end{equation}
the scale in $\phi$ space to which sensitivity has dropped to 50\% is
\begin{equation}
\label{eq:maxscale}
\rm{max~scale} \approx \frac{\pi}{\lambda_{min}^{2}} \sim 143~\rm{rad~m^{-2}}
\end{equation}
and the maximum Faraday depth to which one has more than 50\% sensitivity
\begin{equation}
\label{eq:phimax}
\| \phi_{\rm{max}} \| \approx \frac{\sqrt{3}}{\delta \lambda^{2}} \sim 655~\rm{rad~m^{-2}}.
\end{equation}
The reported numeric values have been obtained for our observations. 
%Comparing Eqns.~\ref{eq:deltaphi} and~\ref{eq:maxscale} we can see where analogy between RM synthesis and regular synthesis imaging breaks down. In synthesis imaging, the width of the synthesized beam is inversely proportional to the maximum absolute uv vector, that is the distance between the origin and the uv point most distant from it. The maximum scale that one can measure depends on the shortest baseline, therefore one is always maximally sensitive to structures smaller than the width of the synthesized beam. In RM synthesis instead it is possible that a source is unresolved in the sense that its extent in $\phi$ is less than the width of the RMSF, yet \enquote{resolved} out because one has not sampled the typical $\phi$-scale of the source due to lack of small $\lambda^{2}$ points. Eq.~\ref{eq:deltaphi} shows that the width of the RMSF depends on the width of the $\lambda^{2}$ distribution, not on the largest $\lambda^{2}$ measured. Nevertheless, the largest scale in $\phi$ that one is sensitive to is set by the smallest $\lambda^{2}$ as is shown in Eq.~\ref{eq:maxscale}.

\begin{figure}
\centering
 \subfloat{\includegraphics[width=\columnwidth]{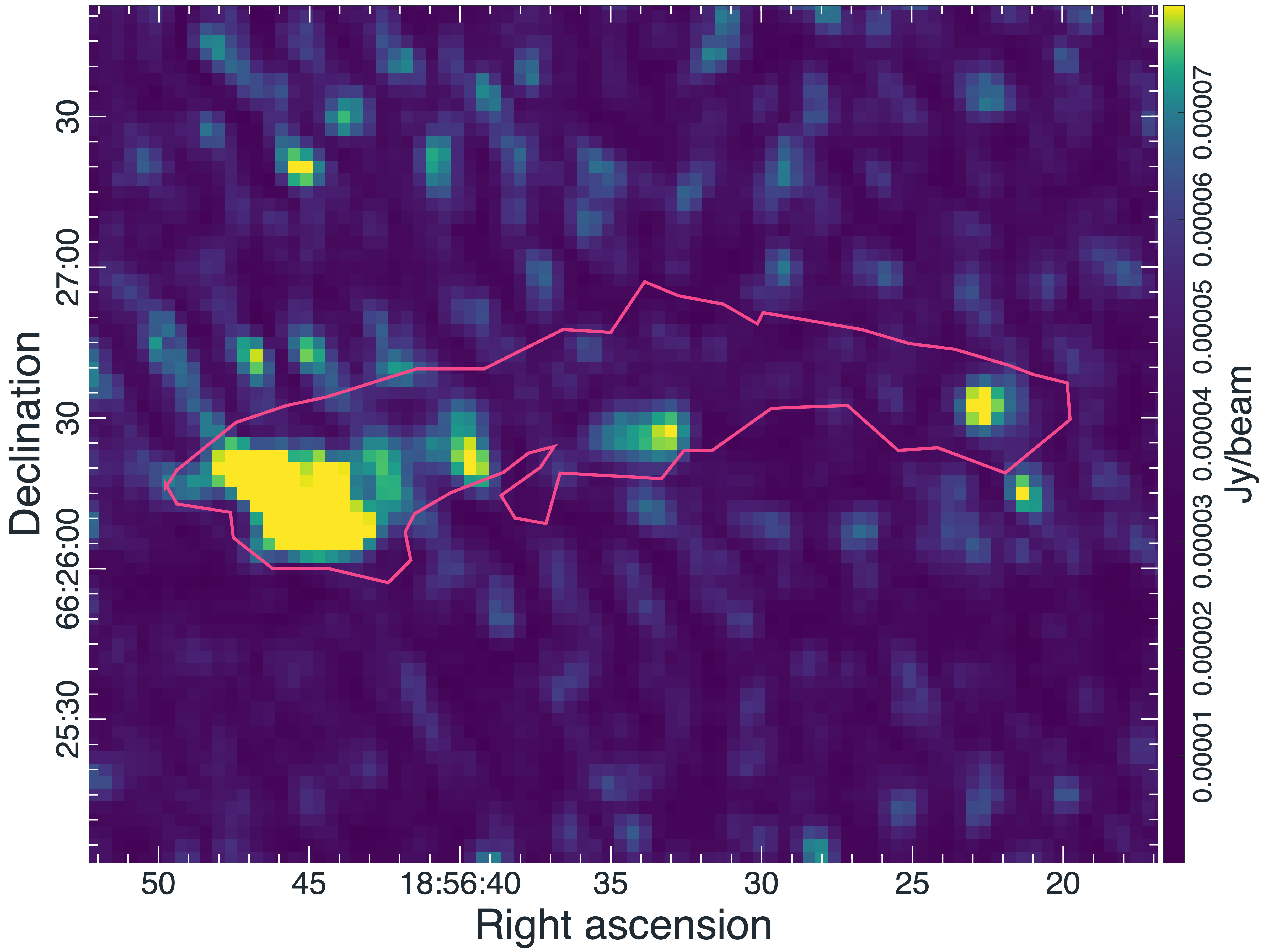}}\\
\subfloat{\includegraphics[width=\columnwidth]{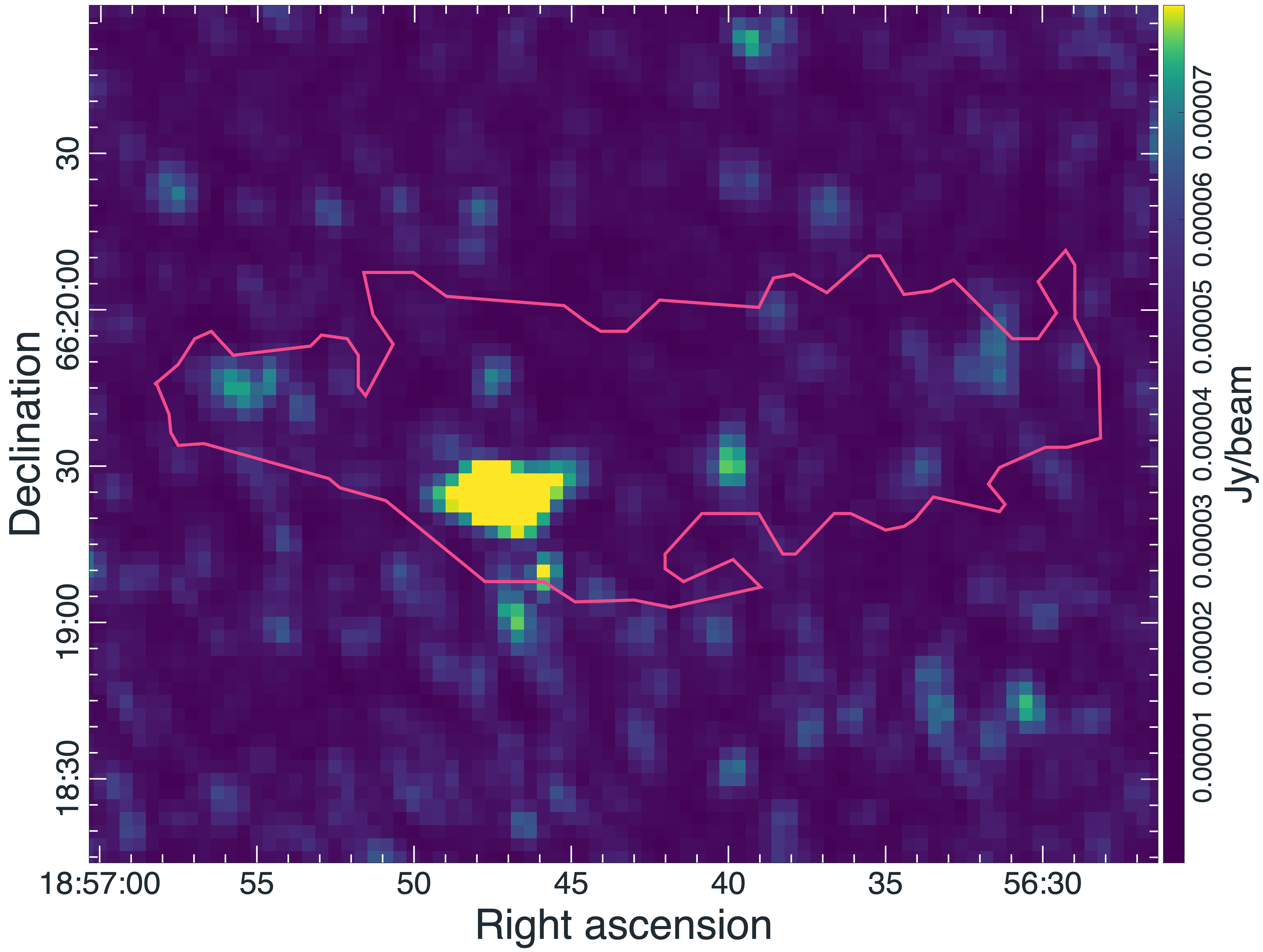}}
\caption{Map of the polarized intensity for each pixel resulting from the RM synthesis, after RM clean (resolution $14" \times 13.2"$, RMS $18~\rm{\mu Jy~beam^{-1}}$). \textit{Top}: northern relic. \textit{Bottom}: southern relic. The two regions chosen for the relics are highlighted in magenta, and correspond to the relics' $5\sigma_{\rm{rms}}$ contour in total intensity.}
\label{fig:PSZ096_relics_regions}
\end{figure}

%%%%%%%%%%%%%%%%%%%%%%%%%%%%%%%%%%%%%%%%%%%%%%%%%%%%%%%%%%%%%%%%%%%%%%%%%%%%%%

% Polarization properties

%%%%%%%%%%%%%%%%%%%%%%%%%%%%%%%%%%%%%%%%%%%%%%%%%%%%%%%%%%%%%%%%%%%%%%%%%%%%%%

%%%%%%%%%%%%%%%%%%%%%%%%%%%%%%%%%%%%%%%%%%%%%%%%%%%%%%%%%%%%%%%%%%%%%%%%%%%%%%%%%%

% RM synthesis results

%%%%%%%%%%%%%%%%%%%%%%%%%%%%%%%%%%%%%%%%%%%%%%%%%%%%%%%%%%%%%%%%%%%%%%%%%%%%%%%%%%

\subsection{Results of RM synthesis}
\label{subsec:mfsrmsynthesisresults}

Beside the integrated analysis of polarization, we applied the RM synthesis technique to resolve the in-band depolarization, and understand if the low polarization fraction observed in the southern relic can be due to bandwidth depolarization, as well as to see if there are complex Faraday effects which can also cause depolarization (like the \enquote{internal depolarization} for example). This technique, as already explained in Sec.~\ref{subsec:RMsynthesis}, uses an observing bandwidth splitted up into many individual narrow frequency channels. Adding up the individual channels may cause bandwidth depolarization: however, using the value of Faraday depth that maximizes the signal resulting from the co-addition of the polarized flux from all channels, it is possible to recover the polarized flux. With the RM synthesis technique it is possible to obtain the reconstructed Faraday spectrum for each pixel. To perform RM synthesis technique on image-frequency cubes we used the \texttt{RMsynth3D} tool present in the \texttt{RM-tools} software. This package takes two input files (Stokes $Q$ and $U$ cubes produced with \texttt{WSClean}) and a list of channel frequencies in order to produce as outputs:

\begin{itemize}
\item The dirty Faraday dispersion function (FDF) cube ($\tilde{F}(\phi)$), which represents the polarized flux as a function of the Faraday depth for each pixel, made by two components: the real (Stokes $Q(\phi)$) and the imaginary (Stokes $U(\phi)$). An example FDF for a single pixel is shown in the bottom panel of Fig~\ref{fig:rmsf_and_fdf}. % Starting from the observed polarized flux, given the instrumental characteristics, the software is able to produce the distribution of the intrinsic polarization with respect to the Faraday depth $\phi$ following Eq.~\ref{eq:reconstructedfdf}.

\item The RMSF (Fig~\ref{fig:rmsf_and_fdf}, top panel), which results from the wavelength coverage and sampling of our observation.%, with the same structure as the dirty FDF. In addition it also contains as the 4-th extension a map of the FWHM of the RMSF.

\item A map of the maximum polarized intensity in each FDF (e.g., for each pixel, Fig~\ref{fig:PSZ096_relics_regions}).

\item A map of the Faraday depth corresponding to the maximum polarized intensity in each FDF.
\end{itemize}
These last two products are calculated from the Faraday depth values sampled by the FDF, without interpolation over the discrete channels. Moreover, we also provided a list of RMS for each frequency-channel of the cube ($\sigma_{\rm{chan}}$, as the average RMS values between $Q$ and $U$ images of each channel) that was used to compute the weight of each frequency channel as the inverse of its RMS, as well as the theoretical noise in P ($\sigma_{\rm{P}}$):
\begin{equation}
\sigma_{\rm{P}} = \frac{1}{\sqrt{\sum_{chan=1}^{56} \frac{1}{\sigma_{\rm{chan}}^{2}}}} = 8~\rm{\mu Jy~beam^{-1}}.
\label{eq:theoreticalnoisep}
\end{equation}
The range of Faraday depths over which we calculate the FDF is set to be $\pm 1000~\rm{rad~m^{-2}}$, while the sample spacing in Faraday depth for the FDF is $3~\rm{rad~m^{-2}}$: the wide range in $\phi$ for the FDF has been chosen in order to use the external intervals to evaluate the noise in each pixel, given that $\|\phi_{\rm max}\|=655~\rm{rad~m^{-2}}$. Once performed the RM synthesis at 13", we also \enquote{cleaned} the dirty FDF using the Faraday dispersion function deconvolution. This deconvolution is possible using the package \texttt{RMclean3D}, that applies the CLEAN algorithm independently to every pixel in a Faraday depth cube. As stopping criterion, a cut-off threshold was set to $8\sigma_{\rm{P}}$, following~\citet{2012george}, which corresponds to a false detection rate of 0.06 \% and to a Gaussian significance level of about 7$\sigma$ according to~\citet{2012hales}. The cut-off threshold was reached for every pixel for which the RM-clean was performed.
\par For the study of the polarization fraction within the two relics, we use the map produced by \texttt{RMsynth3D} showing the maximum polarized intensity in each pixel using cleaned FDF. Then, it is necessary to correct for the Ricean bias (Eq.~\ref{eq:riceanbias}). The value $\sigma_{{P_{\rm{rm,obs}}}}^{2}$ has been computed considering the average RMS noise level between the real and imaginary cleaned FDF evaluated in the ranges $[-1000,-600]~\rm{rad~m^{-2}}$ and $[+600,+1000]~\rm{rad~m^{-2}}$ along the Faraday depth axis (Fig~\ref{fig:rmsf_and_fdf}, bottom panel).
\par With respect to the results obtained with the integrated analysis, owing to the usage of RM synthesis technique, we detected a higher $P$ than in the integrated analysis (Tab.~\ref{tab:fluxes}). Given the flux densities and respective errors, we obtained a value of fractional polarization, $f_{\rm{pol,rm}}$, of
\begin{itemize}
\item $(18.6 \pm 0.3)\%$ for the northern region
\item $(14.6 \pm 0.1)\% $ for the southern region. 
\end{itemize}
These results confirm that bandwidth depolarization has important effects on the polarization fraction of these relics and that we can solve it using the RM synthesis. For the northern region, we obtain a fractional polarization close to the typical value for radio relics~\citep[$\sim 20\%$,][]{2022stuardi}. However, even if the polarization fraction increases with respect to the integrated analysis, the average polarization fraction of the southern relic remains lower than observed in other relics and in the northern one.
\par
\begin{figure}
\subfloat{\includegraphics[width=\columnwidth]{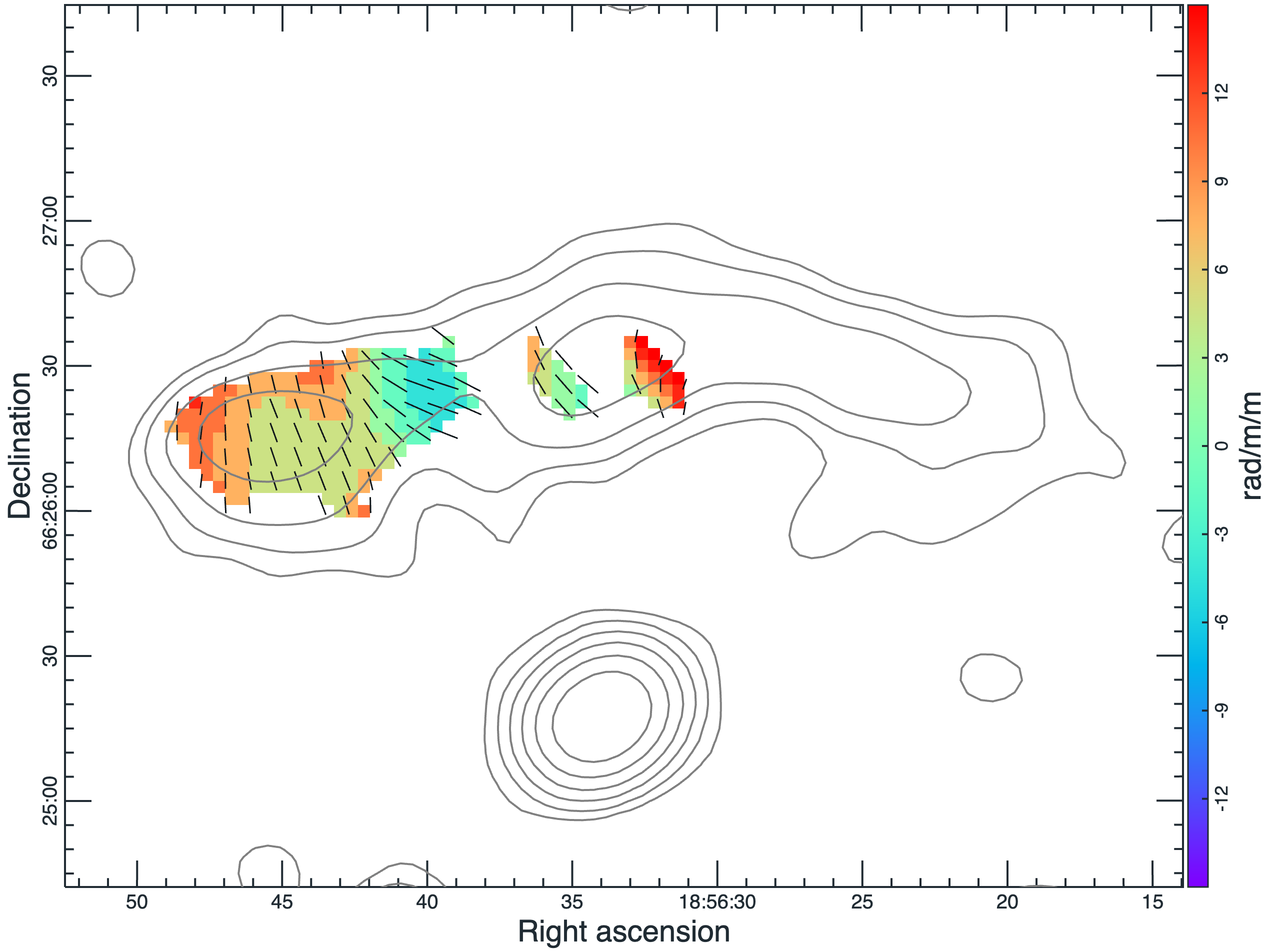}}\\
\subfloat{\includegraphics[width=\columnwidth]{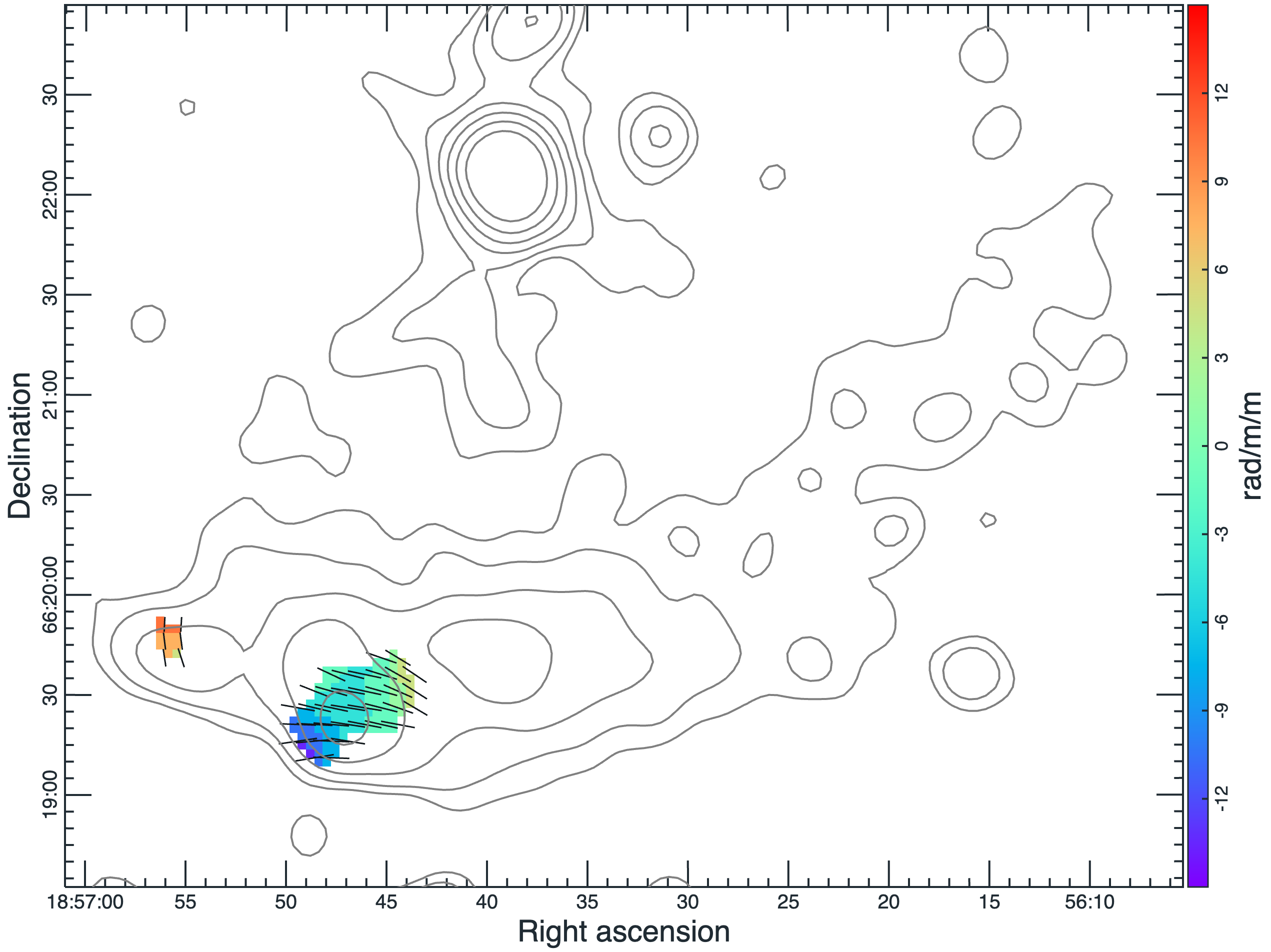}}
\caption{RM map in $\rm{rad~m^{-2}}$ at 30" resolution for the north (top) and south (bottom) relics. Overlaid are the polarization electric field vectors in black, corrected for the galactic RM. The grey contour levels are drawn at $-3\sigma_{\rm{rms}}$, $3\sigma_{\rm{rms}}$, and then increase by a factor of 2, where $\sigma_{\rm{rms}}=15~\rm{\mu Jy~beam^{-1}}$. Only pixels above $6\sigma_{\rm{P}}$ are shown, with $\sigma_{\rm{P}}$ being evaluated pixel by pixel using the external range of their Faraday spectra.}
\label{fig:PSZ096_RM_map}
\end{figure}

\begin{figure}
\includegraphics[scale=0.55]{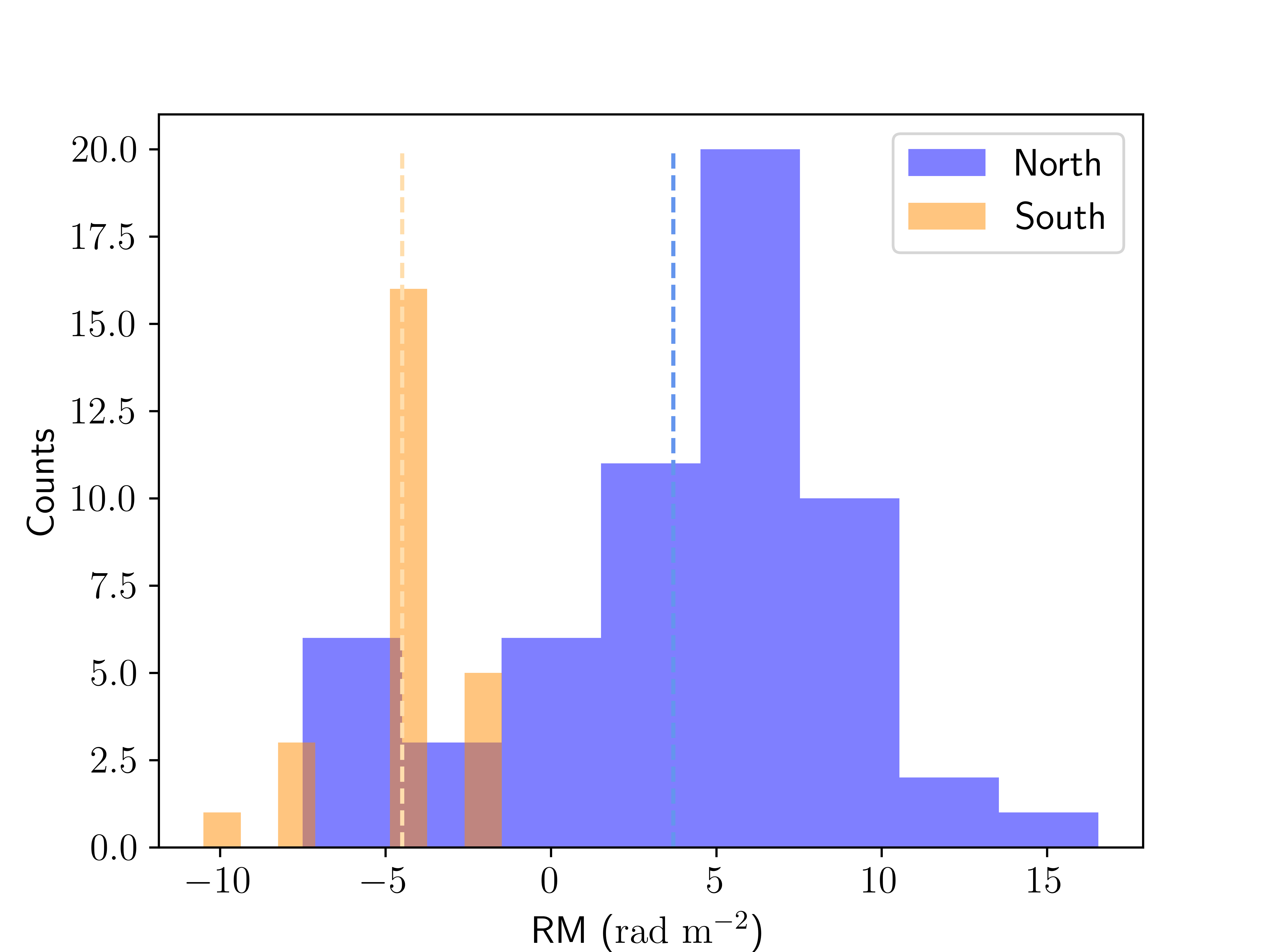}
\caption{Histograms that show the distribution of RM values for the northern (blue) and southern (orange) relic taken from Fig.~\ref{fig:PSZ096_RM_map}. The dashed vertical line represents the average RM values of the two relics. The average $\rm{RM^{2}}$ value is 49.8 for the northern relic and 24.5 for the southern.}
\label{fig:rm_histogram}
\end{figure}

 The RM maps of the two relics are shown in Fig.~\ref{fig:PSZ096_RM_map} and the distribution of RM values over the relics' regions is shown in Fig.~\ref{fig:rm_histogram}. RM values have been corrected for the galactic RM~\citep[$-4.5~\rm{rad~m^{-2}}$, from][]{oppermann2015}. Their low values for both relics suggest that we do not have Faraday rotation within the single channels, and so bandwidth depolarization. The average value of RM is slightly larger for the southern relic ($-4.5~\rm{rad~m^{-2}}$) with respect to the northern one ($3.7~\rm{rad~m^{-2}}$), but the value of $\sigma_{\rm{RM}}$ is higher for the northern relic ($6.6~\rm{rad~m^{-2}}$, against $3.4~\rm{rad~m^{-2}}$ for the southern relic): this indicates that the RM is more uniform in the south relic, while it varies more across the northern relic and this, according to Eq.~\ref{eq:observedpolarization}, should result in more depolarization. In Fig.~\ref{fig:PSZ096_RM_map} we report also the resulting polarization angles, computed using the following Eq.~6 in~\citet{stuardi2019}. We point out that the northern relic has E-field vectors oriented mainly perpendicular to the extent of the relic, as expected for such structures~\citep{2019vanweeren}, with the exception of a few pixels in the upper part, while the southern relic shows E-field vectors oriented parallel to the relic, confirming results in~\citet{2014degasperin}.
\par Another result from RM synthesis is the presence of a small number of pixels that show a complex Faraday spectrum, particularly for the northern relic (see Appendix~\ref{app2} for more details). This implies that our assumption of external Faraday rotation could be too simplistic. After the RM clean, which is based on the theoretical noise level, we notice the presence of some pixels that show multiple components. However, after a careful analysis, we found that most of them are below the noise level, that in these cases is larger that the theoretical prediction. Moreover, they are just a small fraction and are not localized in a restricted area but rather scattered especially at the \enquote{border} of the detection. For all these reasons, we believe that they are not necessarily related to complex behaviors in the emitting regions, and so that the observed RM is mainly external to the relics.
\par The results from RM synthesis and the absence of any significant complex Faraday spectrum at our resolution, let us conclude that the depolarization has an external nature, related to the gas distribution within the cluster in the direction of the relic. %nce resolved the bandwidth depolarization, thanks to the usage of RM synthesis, we will investigate in the next Section the role of beam depolarization in order to understand the true nature of the low fractional polarization that characterizes the southern relic.

%%%%%%%%%%%%%%%%%%%%%%%%%%%%%%%%%%%%%%%%%%%%%%%%%%%%%%%%%%%

% Depolarization trend

%%%%%%%%%%%%%%%%%%%%%%%%%%%%%%%%%%%%%%%%%%%%%%%%%%%%%%%%%%%%

\subsection{Polarization at different resolutions}
\label{subsec:beamdepolarization}

Under the assumption that the depolarization is external and due to the ICM, we want to constrain the magnetic field properties that can reproduce the beam depolarization. To do this, we analyze how the polarization fraction varies with respect to the beam size. The resolution has been changed directly through \texttt{WSClean}, where it has been progressively lowered using a Gaussian taper starting from the 13" to 50" for all the 64 images that constitute the $Q$ and $U$ cubes, as well as the Stokes $I$ MFS images. For each resolution we performed an analysis of polarization fraction following the routine explained in Sections~\ref{subsec:mfsintegratedresults} and~\ref{subsec:mfsrmsynthesisresults}, for both integrated and RM synthesis technique analysis. 
\par
\begin{figure}
\subfloat{\includegraphics[scale=0.55]{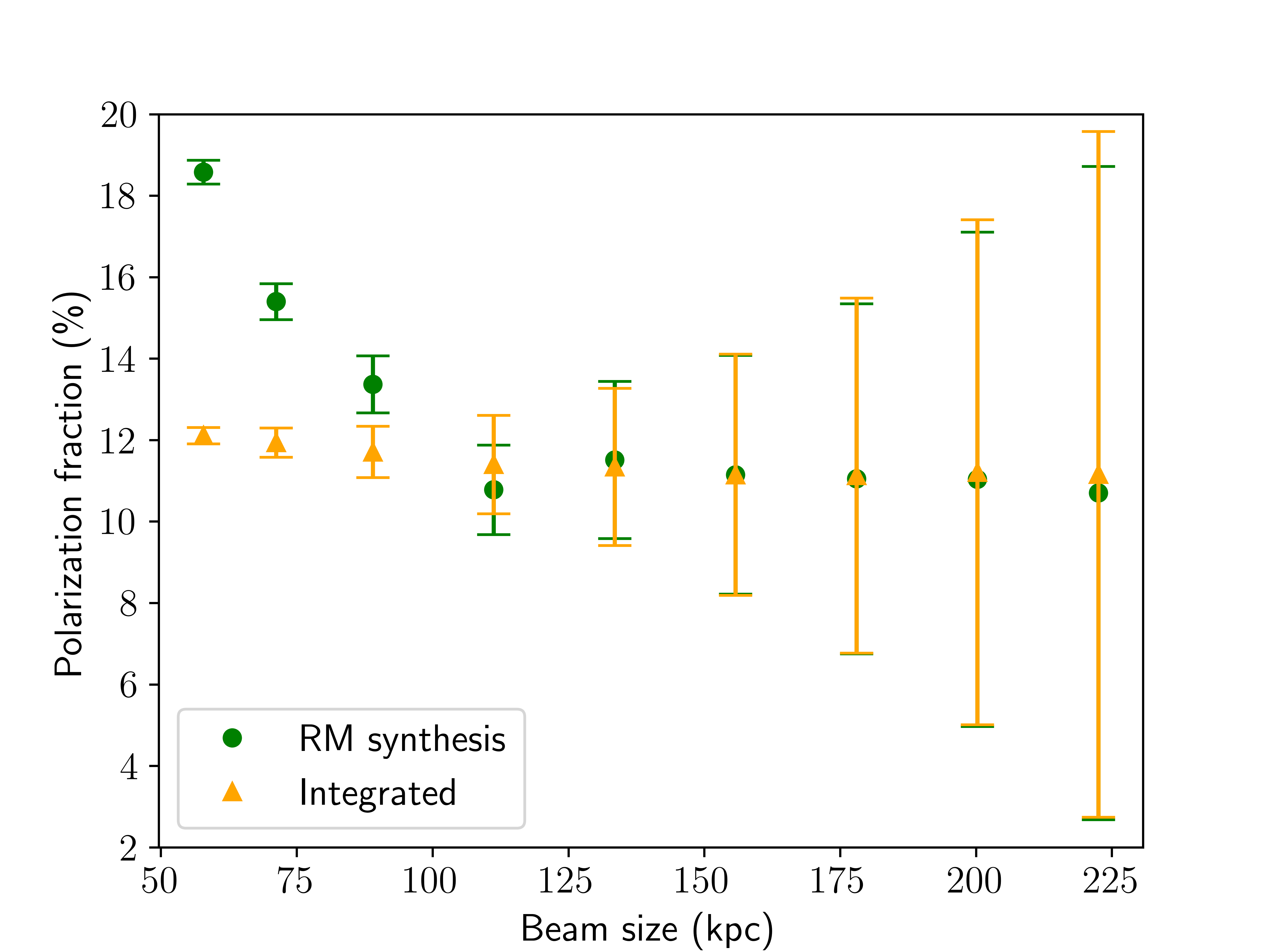}} \\
\subfloat{\includegraphics[scale=0.55]{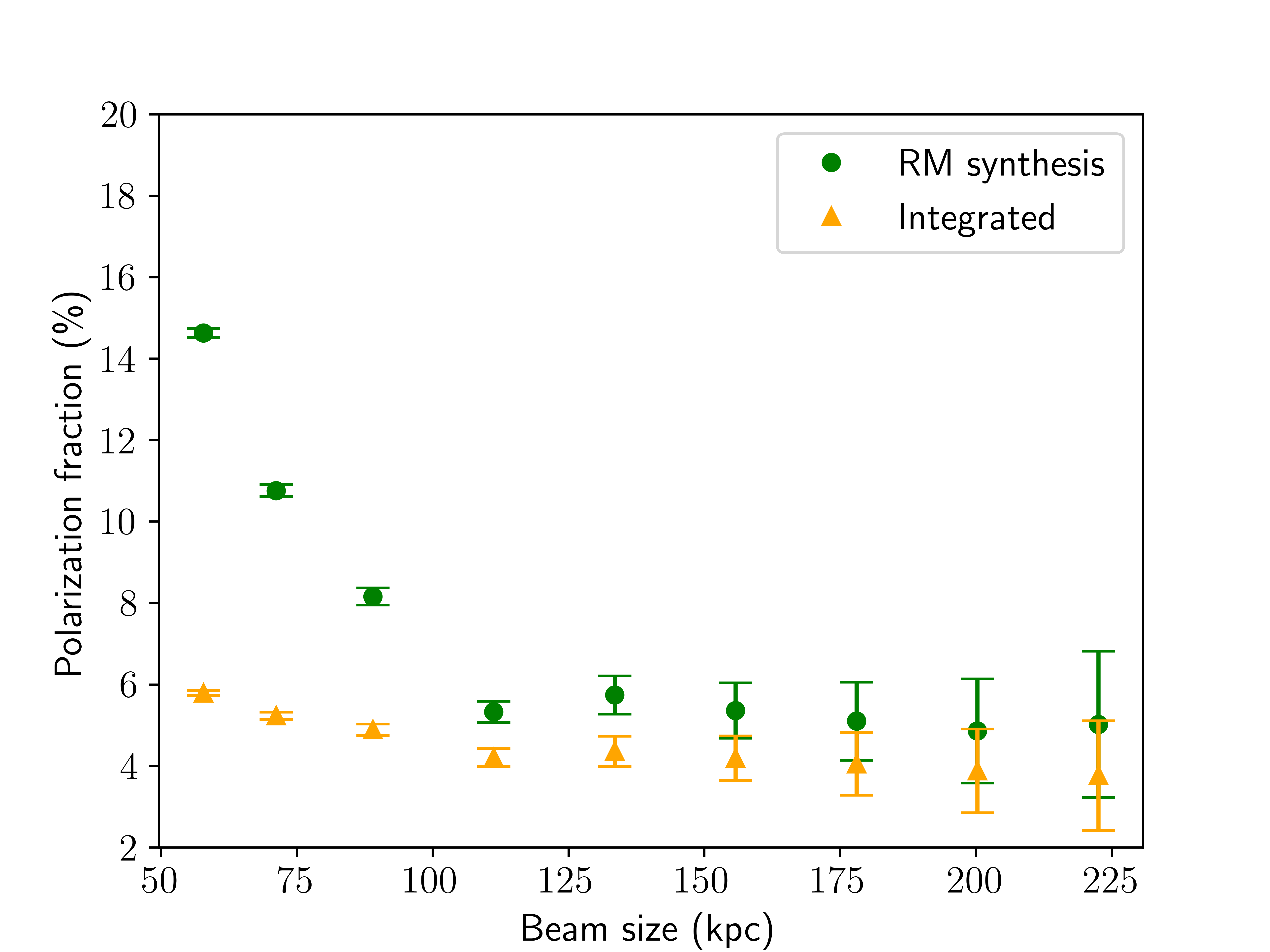}}
\caption{Comparison between the depolarization as a function of beam size evaluated by an integrated analysis (in orange) and using the RM synthesis technique (in green). \textit{Top panel}: for the northern relic. \textit{Bottom panel}: for the southern relic.}
\label{fig:comparison_depolarization_trends}
\end{figure}

 The results for both total intensity ($S_{\rm{I}}$) and polarized ($S_{\rm{P}}$) flux densities, within the regions selected for the relics, are listed in Tab.~\ref{tab:fluxes}, for the integrated analysis and RM synthesis, respectively. Given the flux densities, following Eqs.~\ref{eq:polarizationfraction} and~\ref{eq:polfracerr}, we calculated the trend of polarization fraction as a function of the beam size. The results are listed in Tab.~\ref{tab:fractional_polarizations} for integrated and RM synthesis analysis, respectively. The fractional polarization as a function of the resolution is shown in Fig.~\ref{fig:comparison_depolarization_trends}. 
\par We see that the depolarization decreases with increasing resolution, as expected for beam depolarization. In particular, comparing the two relics, we notice that the decreasing trend is larger for the southern relic: When we evaluate the difference between the fractional polarization at 13" ($\sim$58 kpc) and 25" ($\sim$111 kpc), we find a reduction of the polarization fraction of $(64 \pm 2)\%$ for the southern relic. For the northern relic, instead, we found a decrease of $(42 \pm 5)\%$ in the same range of resolution. A Kolmogorov-Smirnov test shows a statistically significant difference between the depolarization trends of the two relics (KS statistics of 0.78, $p=0.63\%$). The plateau that shows up at lower resolutions (beyond 25", so around 111 kpc in physical scale) points to the fact that the turbulent magnetic field fluctuations within the beam start to become less significant for large beam sizes. We also notice that the error increases for lower resolutions. This is mainly due to the fact that, recalling Eq.~\ref{eq:fluxerror}, we have a progressively smaller number of beams within the regions that have been used to evaluate the flux. At low resolution, the fractional polarization evaluated with RM synthesis is higher than the integrated one for the southern relic, while the two reach approximately the same values for the northern relic. This may indicate higher average RM values also at low resolution for the southern relic. This trend, that depends also on the magnetic field properties, is what we will study in the following part of this work, in order to constrain the characteristics of the cluster magnetic field.

\begin{table}
\centering
\caption{Polarization fractions and respective errors measured for the selected regions for different resolutions.}
\label{tab:fractional_polarizations}
\begin{tabular}{ccc}
\hline\hline
Beam size & $f_{\rm{pol, int}} \pm \delta f_{\rm{pol, int}}$ & $f_{\rm{pol, rm}} \pm \delta f_{\rm{pol, rm}}$ \\
\hline
$13"$ & $(12.1 \pm 0.2)\%$ & $(18.6 \pm 0.3)\%$ \\
$16"$ & $(11.9 \pm 0.4)\%$ & $(15.4 \pm 0.4)\%$ \\
$20"$ & $(11.7 \pm 0.6)\%$ & $(13.4 \pm 0.7)\%$ \\
$25"$ & $(11.4 \pm 1)\%$ & $(10.8 \pm 1)\%$ \\
$30"$ & $(11.3 \pm 2)\%$ & $(11.5 \pm 2)\%$ \\
$35"$ & $(11.2 \pm 3)\%$ & $(11.2 \pm 3)\%$ \\
$40"$ & $(11.1 \pm 4)\%$ & $(11.0 \pm 4)\%$ \\
$45"$ & $(11.2 \pm 6)\%$ & $(11.0 \pm 6)\%$ \\
$50"$ & $(11.2 \pm 8)\%$ & $(10.7 \pm 8)\%$ \\
\hline
\end{tabular}
\begin{tabular}{ccc}
\hline
Beam size & $f_{\rm{pol, int}} \pm \delta f_{\rm{pol, int}}$ & $f_{\rm{pol, rm}} \pm \delta f_{\rm{pol, rm}}$ \\
\hline
$13"$ & $(5.8 \pm 0.1)\%$ & $(14.6 \pm 0.1)\%$ \\
$16"$ & $(5.2 \pm 0.1)\%$ & $(10.8 \pm 0.2)\%$ \\
$20"$ & $(4.9 \pm 0.1)\%$ & $(8.2 \pm 0.2)\%$ \\
$25"$ & $(4.2 \pm 0.2)\%$ & $(5.3 \pm 0.3)\%$ \\
$30"$ & $(4.4 \pm 0.4)\%$ & $(5.7 \pm 0.5)\%$ \\
$35"$ & $(4.2 \pm 0.6)\%$ & $(5.4 \pm 0.7)\%$ \\
$40"$ & $(4.1 \pm 0.8)\%$ & $(5.1 \pm 1)\%$ \\
$45"$ & $(3.9 \pm 1.0)\%$ & $(4.9 \pm 1)\%$ \\
$50"$ & $(3.8 \pm 1.4)\%$ & $(5.0 \pm 2)\%$ \\
\hline
\end{tabular}
\tablefoot{\textit{Top}: results for the northern relic. \textit{Bottom}: results for the southern relic. Column 1: beam size for the different Gaussian kernel. Column 2: integrated polarization fraction and associated error, obtained as seen in Eq.~\ref{eq:polfracerr}. Column 3: same as Column 2 but for RM synthesis.}
\end{table}

%%%%%%%%%%%%%%%%%%%%%%%%%%%%%%%%%%%%%%%%%%%%%%%%%%%%%%%%%%%%

% Simulations presentation

%%%%%%%%%%%%%%%%%%%%%%%%%%%%%%%%%%%%%%%%%%%%%%%%%%%%%%%%%%%%

\section{Comparison to simulations}
\label{sec:simulations}

In this Section, we describe the 3D magnetic field simulations made in order to constrain the magnetic field characteristics of PSZ2G096.88+24.18. We simulated magnetic fields with different power spectra and derived the polarization fraction expected for the southern relic at different resolutions. Then, we compared this resolution dependence with the observational results in order to constrain the best-fit magnetic field parameters that reproduce the observed depolarization trend with the beam size. 

\subsection{Simulation of RM maps}
\label{subsec:simulations of rm maps}

The determination of the cluster magnetic field properties from the RM measurements relies on the knowledge of, both, the thermal electron density and the magnetic field structure (see Eq.~\ref{eq:faradaydepth}). In order to avoid simplistic assumptions, often used to solve the integral for $\phi$, we produced synthetic RM maps by taking into account 3D models of the thermal electron density and of the magnetic field of a galaxy cluster. These RM maps can then be used to compute depolarization trends to be compared with observations, where the magnetic field model parameters can be constrained with a statistical approach. This method has been applied in other work to study cluster magnetic fields, as in ~\citet{2011bonafede} or~\citet{2022osinga}, but, to our knowledge, it has never been applied to the variation of polarization fraction with respect to the resolution. This approach can give us additional information on the physical scales on which the magnetic field fluctuates. In order to simulate a distribution of thermal electron density and magnetic field, we used a modified version of the \texttt{MIR\'O} code described in~\citet{2013bonafede} and in~\citet{2021stuardi}. We present here the different steps.
\par The code starts from a mock 3D thermal electron density distribution of a merging cluster. Due to the complexity of the system and the lack of sensitive observations, it was not possible to use a thermal electron density model derived from X-ray data. Hence, we used a merging galaxy cluster analyzed in~\citet{2019dominguezfernandez} that shares common properties with our observed cluster. It has two density clumps with a projected distance of $\sim 900$ kpc, with a 1:1 mass ratio of the merging sub-clusters. The original simulation had a resolution of 16 kpc at $z=0.02$: it has been then re-scaled in order to have a pixel size of 8 kpc, which is necessary to sample with more than 3 pixels the beam of our highest-resolution observations. This sets the maximum possible wavenumber of the magnetic field power spectrum corresponding to magnetic field fluctuation on the physical scale of 16 kpc. The size of the simulated box is $512^{3}$ pixels ($\sim 4^3~\rm{Mpc^{3}}$), therefore the minimum possible wavenumber corresponds to a physical scale of $\sim2~\rm{Mpc}$.
\par Then, the code produces a 3D distribution of the magnetic field, based on an analytical power spectrum within a fixed range of scales. In this paper, we used both a simplistic Kolmogorov power-law spectrum (where $E_{B} \propto k^{-11/3}$) and a magnetic spectrum derived from~\citet{2019dominguezfernandez} (in this paper we will refer to it as \enquote{PDF})
\begin{equation}
\label{eq:paoladominguezfernandez}
    E_{B}(k)=Ak^{3/2} \Biggl[1-\rm{erf}\Biggl[B \ln{\Biggl(\frac{k}{C}\Biggr)}\Biggr]\Biggr].
\end{equation}
Here $k=\sqrt{\sum_{i} k_{i}^{2}}$ is the wavenumber corresponding to the physical scale of the magnetic field fluctuations (e.g., $\Lambda \propto 1/k$), with $i$ representing its three components, the $A$ parameter gives the normalization of the magnetic spectrum, $B$ is related to the width of the spectra, and $C$ is a characteristic wavenumber corresponding to the inverse outer scale of the magnetic field. The parameters $B$ and $C$ were taken from the E5A simulated cluster in~\citet{2019dominguezfernandez}\footnote{\label{note4} see Tab.1 of~\citet{2019dominguezfernandez}}, having values of $B = 1.054$ and $C = 8.708~k[1/2~\rm{Mpc}]$.
\par In order to obtain a divergence-free turbulent magnetic field with this power spectrum, we first selected the corresponding power spectrum for the vector potential $\tilde{\textbf{A}}(\textbf{k})$ in Fourier space~\citep{2004murgia, 1991tribble}. For each pixel, in Fourier space, the amplitude, $A_{k,i}$ (where $i$ represent its three components), is taken from a Rayleigh distribution in order to have magnetic field components that follow a Gaussian distribution, and the phase of each component of $\tilde{\textbf{A}}(\textbf{k})$ is randomly drawn in range [0,$2\pi$]. The magnetic field vector in Fourier space is then $\tilde{\textbf{B}}(\textbf{k})=i\textbf{k} \times \tilde{\textbf{A}}(\textbf{k})$ and has the desired power spectrum. $\tilde{\textbf{B}}(\textbf{k})$ is transformed back into real space using an inverse Fast Fourier Transform (FFT) algorithm. The resulting magnetic field, \textbf{B}, has components $B_{i}$ following a Gaussian distribution, with $\langle B_{i} \rangle =0$ and $\sigma_{B_{I}}^{2}=\langle B_{i}^{2} \rangle $.
\par The radial profile of the magnitude of the magnetic field is expected to scale with the thermal electron density. This radial decrease of the magnetic field strength is also expected by MHD simulations~\citep[e.g.,][]{2018vazza, 2019dominguezfernandez}. Therefore, we imposed that the cluster magnetic field scales with the thermal electron density following a power-law
\begin{equation}
\|B(r)\| \propto n_{e}(r)^{\eta},
\end{equation}
where $\eta$ is set at 0.5, as obtained in~\citet{2010bonafede}, which we expect if the energy in the magnetic field scales as the energy in the thermal plasma (assumed to be isothermal).
\par The magnetic field is scaled by the density profile and then normalized for a value $B_{\rm{mean}}$ over the entire box. Hence, the generated cluster magnetic field is tangled on both small and large scales, and it decreases with the thermal electron density. In the case of the PDF power spectrum, $B_{\rm{mean}}$ set also the $A$ parameter of the analytical formula in Eq.~\ref{eq:paoladominguezfernandez}. $B_{\rm{mean}}$ has been chosen in order to give, at the southern relic position, a magnetic field equal to the one evaluated with the equipartition assumptions. Following Eq. 26 in~\citet{2004govoni}, we find an equipartition magnetic field of $0.67~\rm{\mu G}$\footnote{\label{note5} assuming a depth intermediate between length and height, a radio brightness at of $S_{\rm{1.5~GHz}} = 7.37 \times 10^{-4}~\rm{mJy~arcsec^{-2}}$ with an $\alpha = 1.17$~\citep[values taken from][]{2021jones}, $\rm{k}=1$ as parameter describing the nature of the plasma (typical value for galaxy clusters) and $\xi(\alpha,\nu_{1},\nu_{2}) = 3.42 \times 10^{-13}$~\citep[as listed in Tab. 1 in][]{2004govoni}}. The value of $B_{\rm{mean}}$ that gives us the equipartition value at the southern relic position is $B_{\rm{mean}}=0.8~\rm{\mu G}$, so we decided to fix this value in order to limit the number of free parameters for the simulations.
\par Finally, the code computes the cluster 2D RM map integrating the thermal electron density and magnetic field profile along one axis, solving Eq.~\ref{eq:faradaydepth}, from the centre of the cluster, thus assuming that the sources lie on the plane parallel to the plane of the sky and crossing the cluster center.
\par To summarize, we fixed the following parameters for the simulations: the size and resolution of the box, the parameter $\eta$, the factors $B$ and $C$ of the PDF magnetic field power spectrum and the normalization factor $B_{\rm{mean}}$. The only free parameters are the minimum and maximum scale over which the magnetic field power spectrum is defined.

\subsection{Simulated polarization fraction}

For the simulation of the polarization fraction at different resolutions we used the 2D RM maps produced by the \texttt{MIR\'O} code. To compute the depolarization effect for different magnetic field models we used an approach similar to the one used in~\citet{2022osinga}\footnote{\label{note3} \url{https://github.com/ErikOsinga/magneticfields}.}. For the simulated fractional polarization, we have to assume an intrinsic polarization vector for each pixel in our simulated RM map. We assumed a uniform distribution with $\chi_{0} = 45\degr$. The choice of the intrinsic polarization angle is arbitrary, while the uniform distribution is driven by the assumption that intrinsic polarization vectors are all aligned by the shock passage. We also assumed an intrinsic polarized flux density of $P_{\rm{intr}} = 1~\rm{Jy~beam^{-1}}$ for each pixel. This is an arbitrary choice since we are only interested in the change of polarization with the beam size, $P/P_{\rm{intr}}$, where $P$ is the polarized flux measured after smoothing.
\par
\begin{figure}
 \includegraphics[width=1.1\columnwidth]{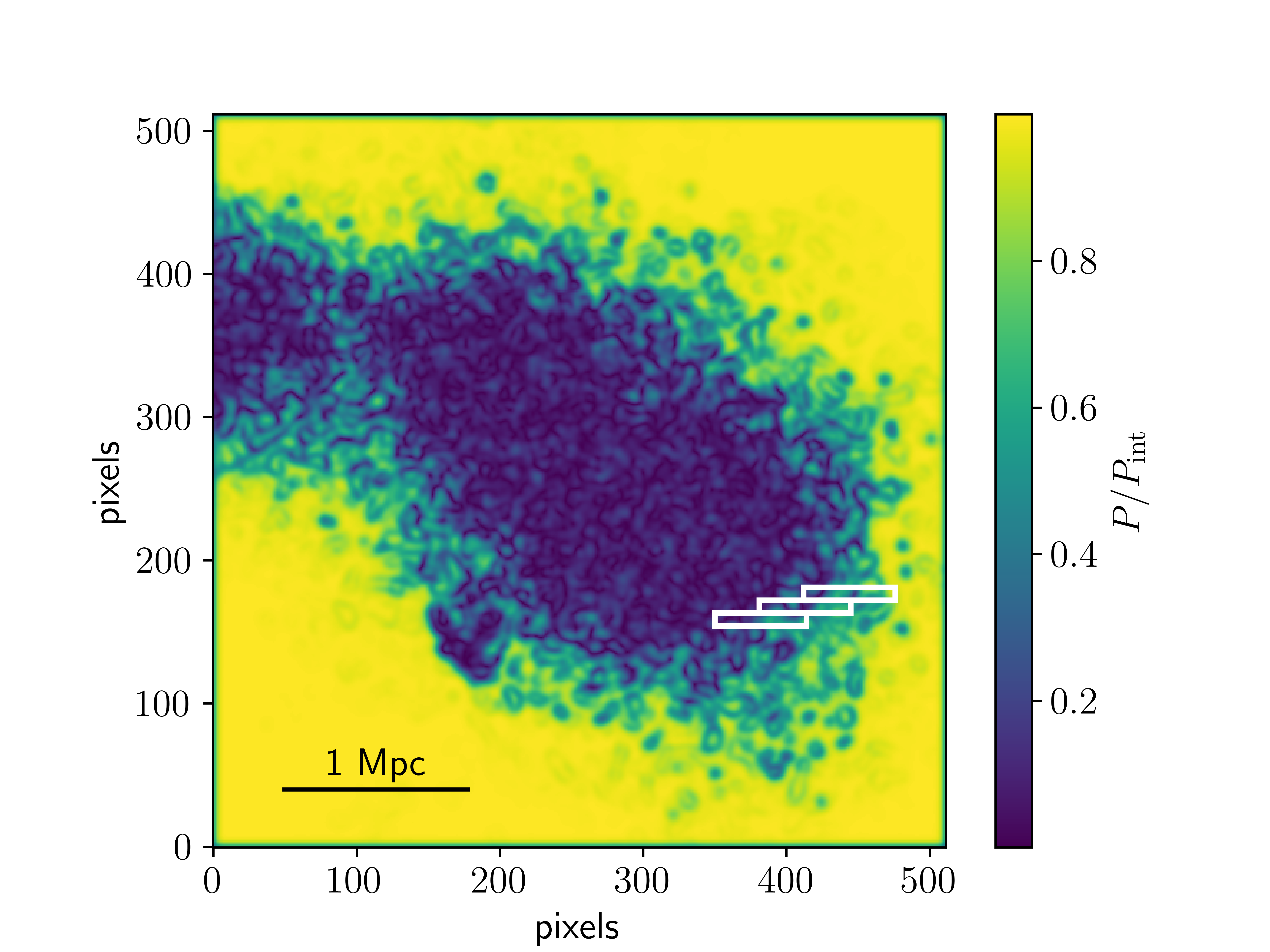}
 \caption{2D smoothed map at $13\arcsec$ of polarized intensity over intrinsic polarized intensity, resulting from depolarization code presented in~\citet{2022osinga}. In white we highlight the region chosen to extract the simulated polarization fraction for the southern relic.}
 \label{fig:PSZ096_sim}
\end{figure}

 Once we have obtained $\chi$ with Faraday rotation using the simulated RM map ($\chi = \chi_{0} + \lambda^{2} \rm{RM}$), and using $P$ we can compute the Stokes $Q$ and $U$ parameters using the following relations
\begin{equation}
\label{eq:pol_stokes}
P=\sqrt{Q^{2}+U^{2}} \quad,\quad \chi=\frac{1}{2}\arctan{\frac{U}{Q}}~,
\end{equation}
which we can convert to
\begin{equation}
\label{eq:osinga}
Q=\pm \sqrt{\frac{P^{2}}{1+\tan^{2}{(2\chi)}}} \quad,\quad U=\pm \sqrt{P^{2}-Q^{2}},
\end{equation}
using the convention that Stokes $Q$ is positive for $-\pi/2 \leq \chi \leq \pi/2$ and Stokes $U$ is positive for $0 \leq \chi \leq \pi$. The $Q$ and $U$ simulated images have been smoothed with a Gaussian kernel with dimensions equal to the ones used to trace the observed depolarization trend, and the smoothed $P$ is computed again using Eq.~\ref{eq:pol_stokes}. Finally, we generated the maps of $P/P_{\rm{intr}}$, at different resolutions (an example is shown in Fig.~\ref{fig:PSZ096_sim}).
\par From the simulations we obtain $P/P_{\rm{intr}}$, which is the loss of polarization intensity due to Faraday rotation at different resolution and not the polarization fraction that is what we measure from observations. Thus we need to re-scale it to compare data and model:
\begin{equation}
\frac{P}{P_{\rm{intr}}} = \frac{P/I_{\rm{intr}}}{P_{\rm{intr}}/I_{\rm{intr}}} = \frac{f_{\rm{pol,sim}}}{f_{\rm{pol,intr}}}~.
\end{equation}
This means that we have a relationship between the simulated and the intrinsic polarization fraction which is:
\begin{equation}
\label{eq:fracpolsim_vs_int}
f_{\rm{pol,sim}} = \frac{P}{P_{\rm{intr}}} \cdot f_{\rm{pol,intr}}.
\end{equation}
Given that we do not know the intrinsic polarization fraction we used an assumption based on~\citet{1998enblin}. According to their formalism, if the magnetic pressure of the relics is small compared to the internal gas pressure the compression of the magnetized regions is equal to the compression of the accretion shocks. If we use the values related to the southern relic\footnote{$\alpha = 0.97$~\citep[$\gamma=2\alpha+1$;][]{2021jones}, $\delta = 90\degr$, $R = (\alpha + 1)/(\alpha - 1/2)$.}, gives a $f_{\rm{pol,intr}} \sim 54.7\%$~\citep[using Eq. 22 in ][]{1998enblin}. We assume this as the intrinsic polarization fraction $f_{\rm{pol,int}}$. This means that we have to scale our simulation by this value, given the Eq.~\ref{eq:fracpolsim_vs_int}, in order to obtain the simulated polarization fraction, that we need to compare data and models. We selected a region in the simulated polarization maps (highlighted as white boxes in Fig.{\ref{fig:PSZ096_sim}}) with the same position and dimension of the southern relic~\citep[as reported in ][]{2021finner, 2021jones}. Within this region, we evaluated the polarization fraction for the southern relic at different resolutions, matching those in our observations.

\subsection{Data and simulations comparison}
\label{sec:data_sim_comparison}

To compare data and simulations, and assess which magnetic field power spectrum better reproduces the observed depolarization trend, we use a reduced chi-squared test, that takes into account the error on the observed polarization fraction, as well as the observed and simulated ones. To consider the statistical fluctuations related to the simulations, we made a set of 10 runs for each magnetic field power spectrum, with a certain combination of minimum and maximum scales. Then, we averaged the values obtained at each resolution in order to get an average trend, taking into account all the simulation runs, and this average trend is then compared with the observed data using the reduced chi-squared test
\begin{equation}
    \chi^{2} = \sum_{i=1}^{9} \frac{(O_{i}-\langle C_{i}\rangle_{10})^{2}}{{\rm{scatter_{sim}}}^{2}+\delta f_{{\rm{pol,rm}}~i}^{2}}.
    \label{eq:reduced_chisquared}
\end{equation}
This evaluation of the reduced chi-squared comes from~\citet{govoni2006}, where $i$ represents the nine resolutions used for the beam size, $O_{i}$ are the observed data, $\langle C_{i}\rangle_{10}$ represent the simulated data averaged over the 10 identical run of simulations for any combination of scales,  $\delta f_{{\rm{pol,rm}}~i}$ represents the error on the observed data, and $\rm{scatter_{sim}}$ represents the standard deviation of the simulations.
\par Now we only vary the scales $\Lambda_{\rm{min}}$ and $\Lambda_{\rm{max}}$ over which the magnetic field power spectrum is defined. The minimum scale of 16 kpc is limited by our observational capabilities, while for the maximum scale we are limited by the size of the problem, because we wanted to execute and replicate the simulations multiple times within acceptable run-times. We expect $\Lambda_{\rm{min}}$ below 16 kpc not only from theory~\citep{2014brunetti}, but also from the fact that at 13" resolution we already observe depolarization with respect to the expected intrinsic one for the relic~\citep[$\sim 54.7\%$ from][]{1998enblin}. On the contrary, given the plateau observed over 100 kpc in Fig.~\ref{fig:comparison_depolarization_trends}, we do not expect significantly higher $\Lambda_{\rm{max}}$ otherwise we should have observed additional depolarization towards larger beam sizes. Once defined these boundaries for the possible minimum ad maximum scales, we performed multiple simulations varying the combination of scales.
\par
\begin{figure}
\subfloat{\includegraphics[width=\columnwidth]{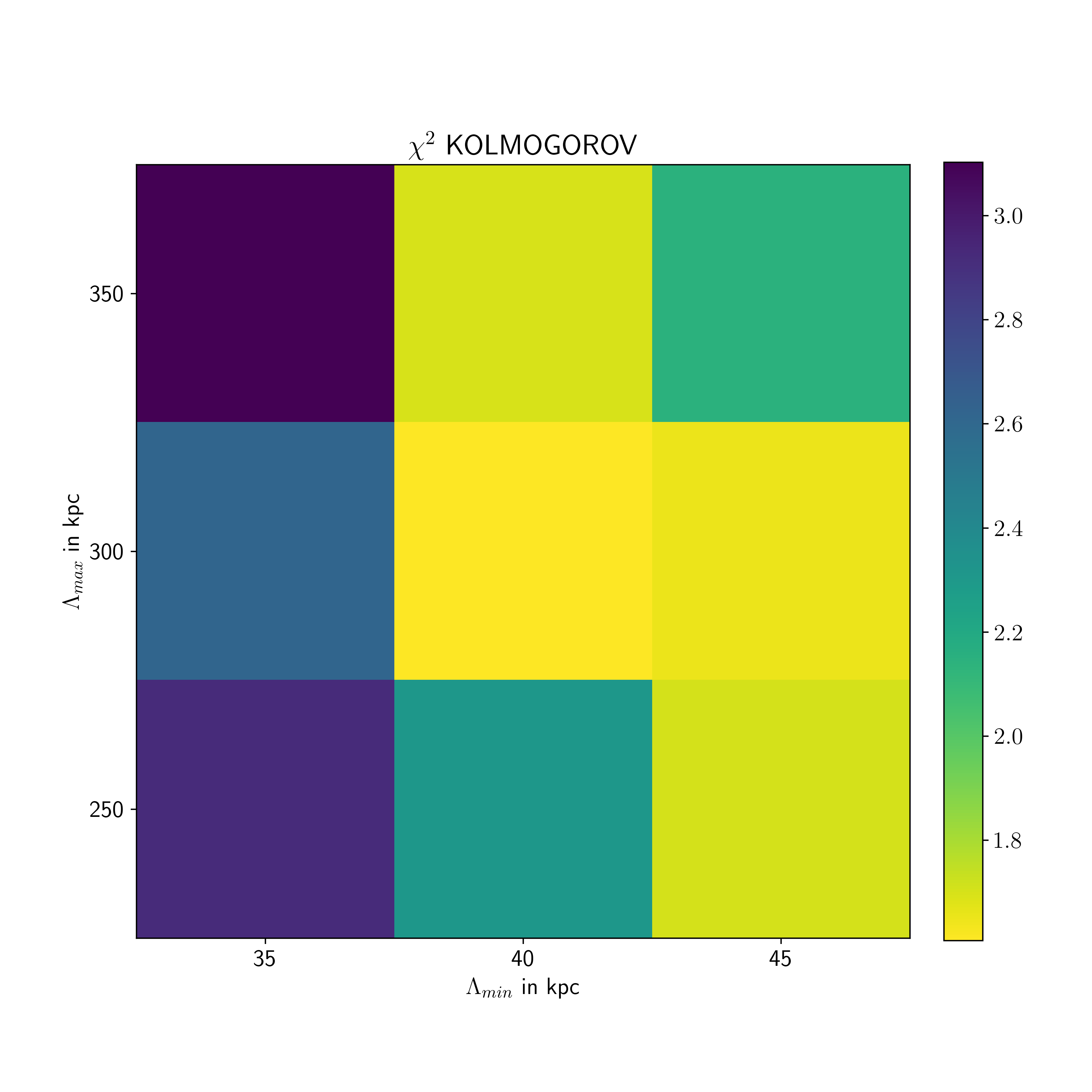}}\\
\subfloat{\includegraphics[width=\columnwidth]{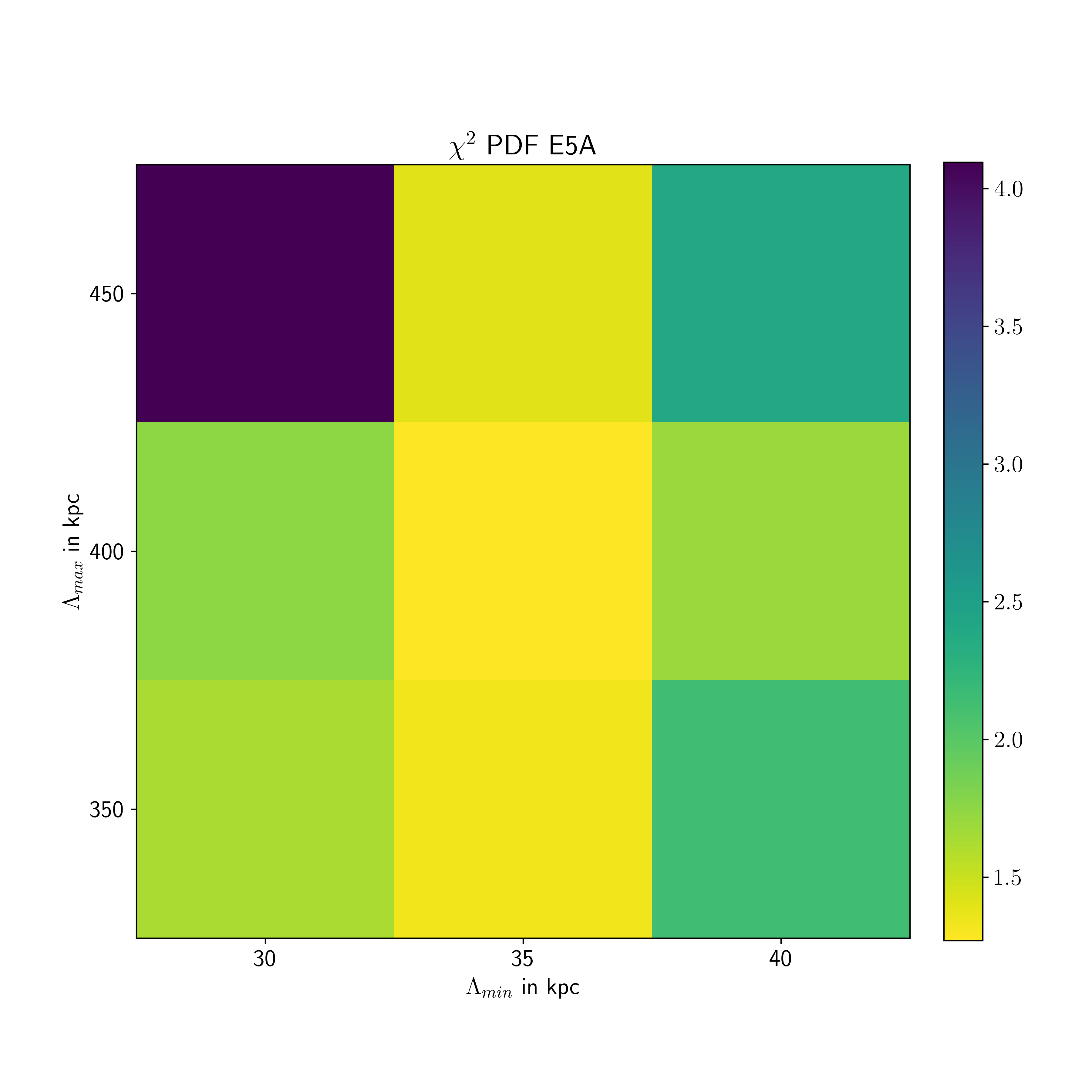}}
\caption{Comparison between the reduced chi-squared values obtained for the two magnetic field power spectra selected for this work. The two axes represent the minimum and maximum scales in kpc over which the spectra are defined. In color scale we have the reduced chi-squared evaluated following Eq.~\ref{eq:reduced_chisquared}. \textit{Top panel}: for the Kolmogorov power spectrum. \textit{Bottom panel}: for the PDF power spectrum.}
\label{fig:comparison_chisquared_trends}
\end{figure}

 The results of the comparison between data and simulations are represented in Fig.~\ref{fig:comparison_chisquared_trends} where we show, for each combination of minimum and maximum scales, the value of the reduced chi-squared in different color scale. It is important to underline that the ones in the Figure represent just a small sub-sample of all the scales that we used: we just represent these because they surround the best-fit combination of scales. For the Kolmogorov magnetic field power spectrum, we find that the combination of minimum and maximum scales that minimize the reduce chi-squared is $\Lambda_{\rm{min}}=40~\rm{kpc}$ and $\Lambda_{\rm{max}}=300~\rm{kpc}$: this results in a $\chi_{\rm{min}}^{2}=1.61$. For PDF power spectrum, instead, we find the best fit for a minimum scale of $\Lambda_{\rm{min}}=35~\rm{kpc}$ and a maximum scale of $\Lambda_{\rm{max}}=400~\rm{kpc}$: this gives a reduced chi-squared value of $\chi_{\rm{min}}^{2}=1.27$.
\par
\begin{figure}
\subfloat{\includegraphics[width=\columnwidth]{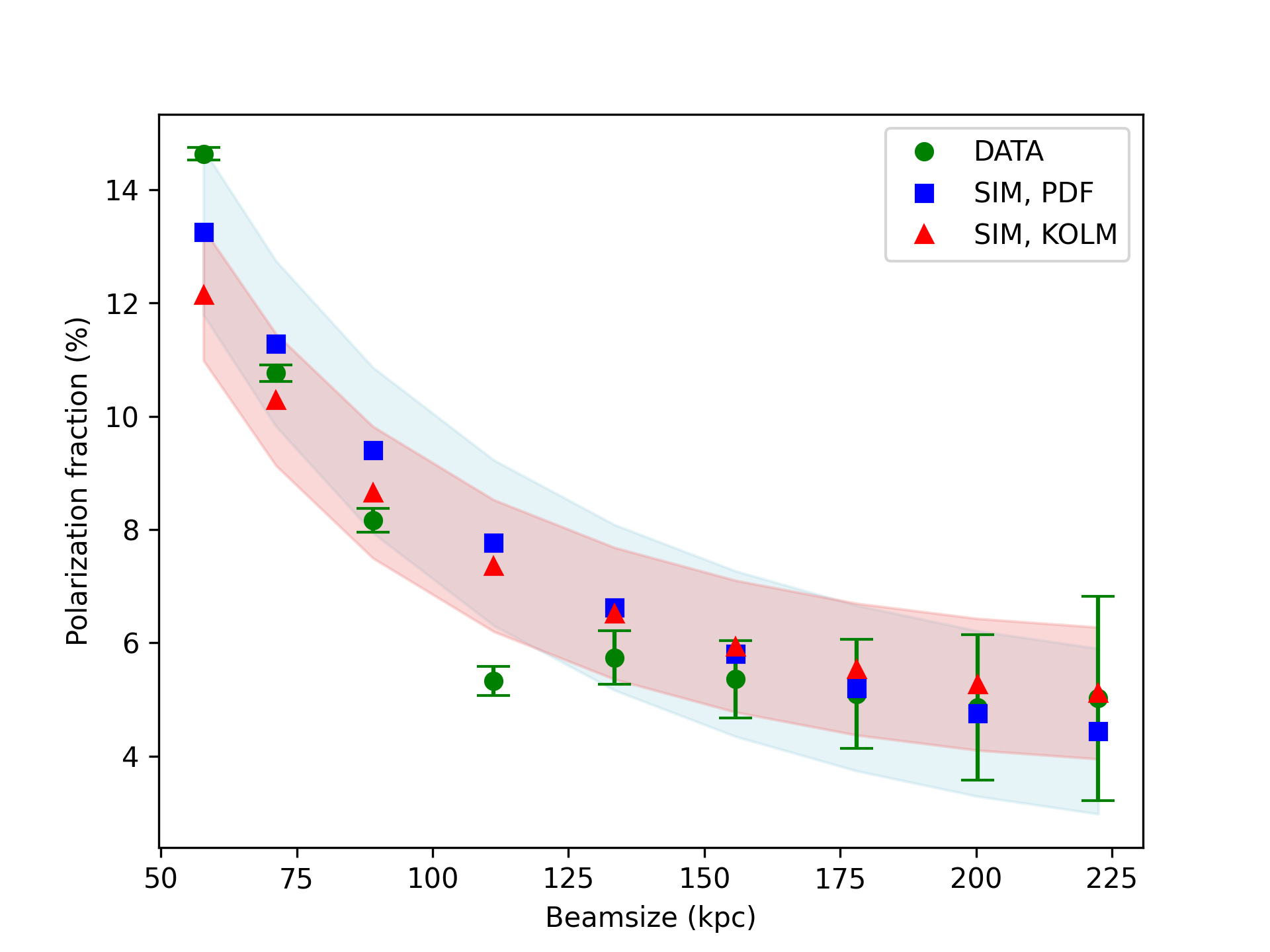}}\\
\subfloat{\includegraphics[width=\columnwidth]{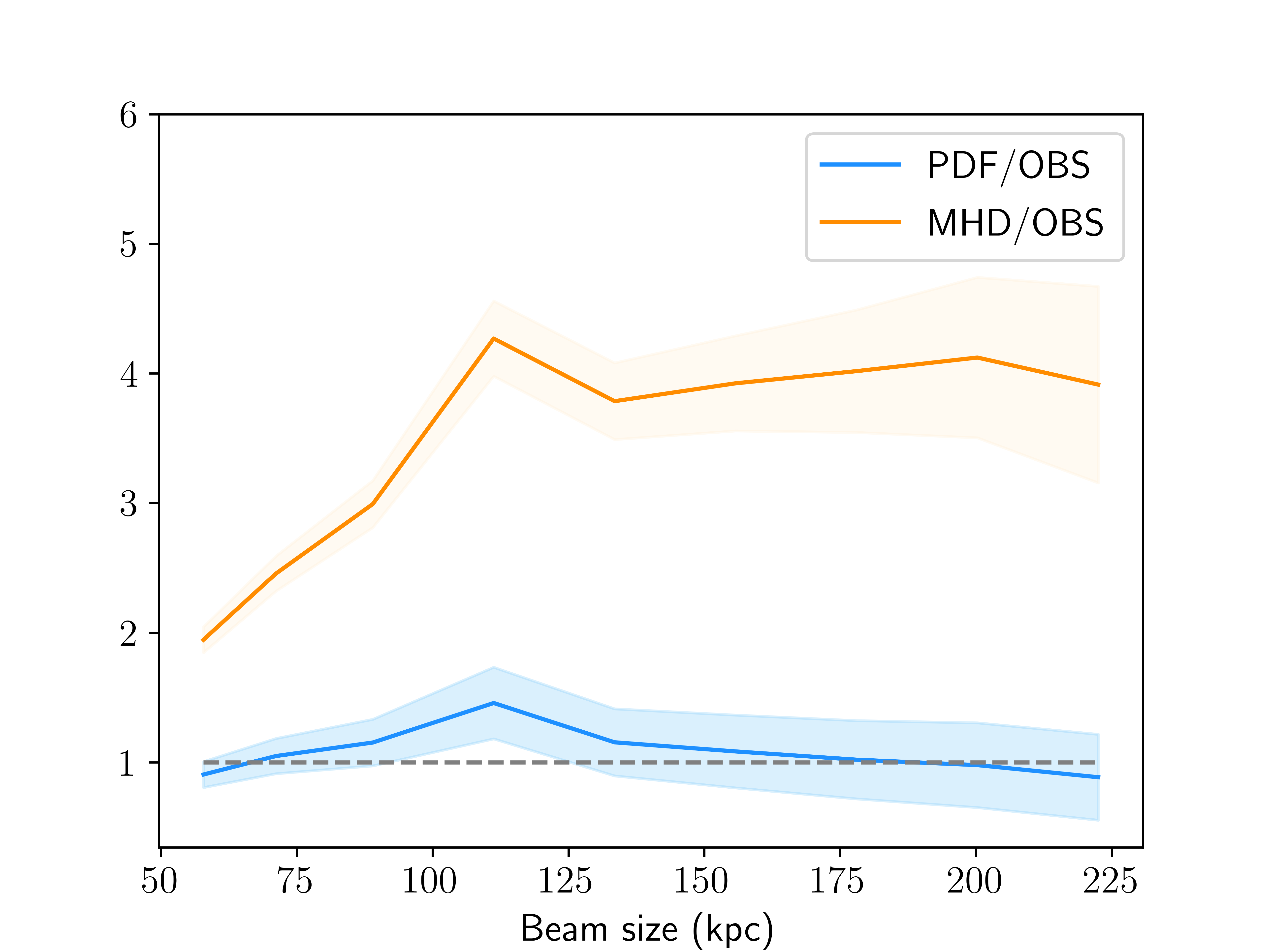}}
\caption{\textit{Top panel}: observed depolarization trend with the beam size (green circles) compared with the best fit models for both the Kolmogorov (red triangles) and PDF (blue squares) magnetic field power spectrum (Sec.~\ref{sec:data_sim_comparison})\\
\textit{Bottom panel}: comparison between the simulated depolarization trends normalized for the observed one. The best fit PDF magnetic field power spectrum found in Sec.~\ref{sec:data_sim_comparison} is in blue, while the magnetic field from MHD simulations is in orange. The dashed, horizontal grey line represents the perfect coincidence between simulation and observational trends.}
\label{fig:PSZ096_obs_vs_sims}
\end{figure}

 Together with the magnetic field generated by \texttt{MIR\'O}, we also performed the same analysis using the three-dimensional magnetic field directly originated by the MHD simulations in~\citet[][E5A cluster]{2019dominguezfernandez}. The resulting depolarization trend does not reproduce the observed one: using the magnetic field from the MHD simulation we obtain a depolarization trend with a higher normalization and a shallower reduction of polarization fraction with increasing beam size (Fig.~\ref{fig:PSZ096_obs_vs_sims}). This results in a reduced chi-squared of $\chi^{2}_{\rm{MHD}} = 245$. Comparing the average magnetic field computed at the relic position, we find a value for the MHD simulations of a factor $\sim 3$ smaller with respect to the one we imposed with \texttt{MIR\'O} to match the equipartition value (Sec.~\ref{sec:simulations}). The higher polarization fraction is thus consistent with the lower magnetic field intensity of the MHD field. The different depolarization trend suggests that the MHD field has a more complex distribution with respect to the one generated by \texttt{MIR\'O}, in which we have also increased the resolution, introducing fluctuation on smaller scales.

%_____________________________________________________________________%
%   DISCUSSION
%_____________________________________________________________________%

\section{Discussion}
\label{sec:discussion}

%\begin{figure}
%\subfloat{\includegraphics[width=\columnwidth]{statisticsPSZ096_KOLMOGOROV_PAPER_sim_bmean08_40min_300max.pdf}}\\
%\subfloat{\includegraphics[width=\columnwidth]{statisticsPSZ096_PDF_PAPER_sim_bmean08_35min_400max.pdf}}
%\caption{Comparison between the simulated magnetic power spectrum generated for the two models used in this paper (\textit{Top panel}: for the Kolmogorov power spectrum. %\textit{Bottom panel}: for PDF power spectrum). On the x-axis, $k$ is represented as $(4096~\rm{kpc})/\Lambda$. The scales chosen for these plots correspond to the best fit scales found for each model (see Sec.~\ref{sec:data_sim_comparison}).}
%\label{fig:comparison_power_spectrum}
%\end{figure}

\subsection{Origin of the low polarization fraction of the southern relic}

In this work, we wanted to study the nature of the low polarization fraction observed in the southern relic of PSZ2 G096.88+24.18 compared to the northern relic using RM synthesis. The origin of the different behavior of polarization in the two relics is not yet understood, and has been investigated in previous work.
\par \citet{2014degasperin} suggested that this could be caused by the southern relic lying further away from us compared to the northern relic. In this scenario, the observed radio emission would have to pass through more magnetized, ionized plasma and therefore be subject to more Faraday rotation. Spectroscopic observations of PSZ2 G096.88+24.18 show that the redshift distribution of the member galaxies is well fit by a single Gaussian~\citep{2019golovich}. This makes it unlikely that significant additional Faraday rotation due to projection effects is causing the observed difference in polarization angle. However, we need to consider that gas and galaxies can have a different distribution.
\par
\begin{figure}
\includegraphics[width=\columnwidth]{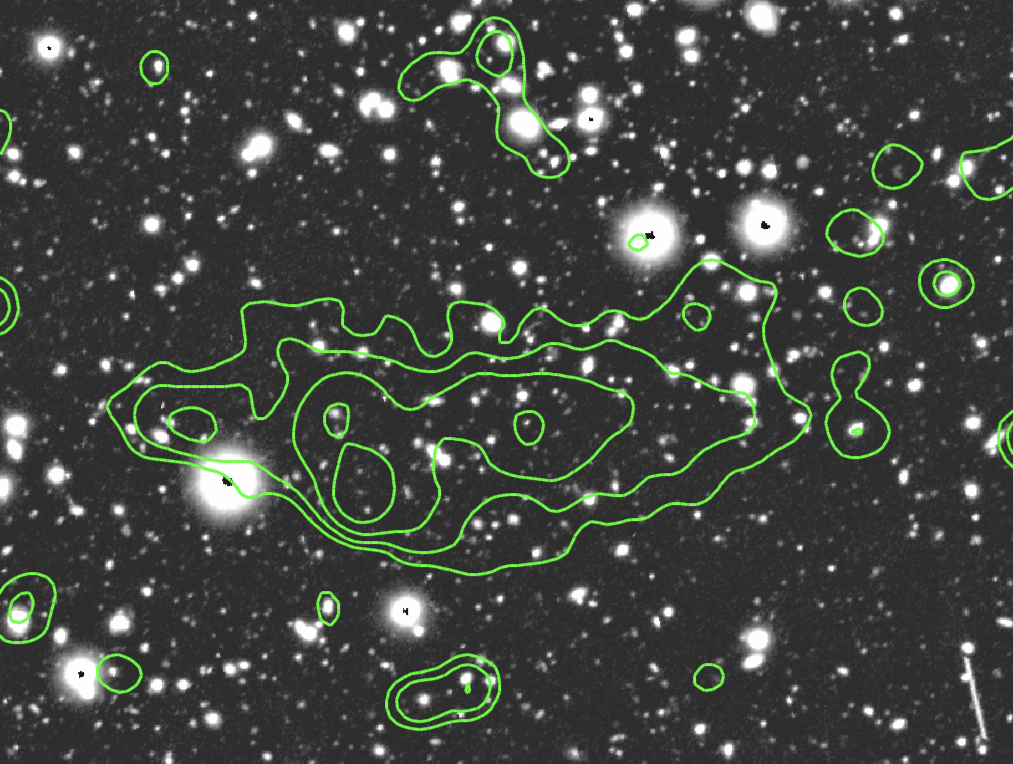}
\caption{Total intensity radio contours (same as in Fig.~\ref{fig:PSZ096_totalintensity_10asec} overlayed on a Subaru/Suprime-Cam image of the cluster. We notice that, corresponding to one of the brightest part of the southern relic, we have possible contaminating background sources which can contribute to the relic polarization.}
\label{fig:subaru_contours}
\end{figure}

 The localized nature of the polarized emission and unexpected electric field vector orientation could instead indicate that we are observing emission from a polarized radio galaxy in projection within the relic.~\citet{2021jones} claimed that no optical counterpart was observed and the polarized emission coincides with the brightest part of the relic. However, comparing our radio data with deep optical image (Fig.~\ref{fig:subaru_contours}), we notice the presence of three possible contaminating sources, which can contribute to the polarization of the southern relic. Therefore, we cannot exclude the contribution of background sources, even if given their generally low polarization fraction (below $10\%$) and the fact that we evaluate the fractional polarization using flux density taken over the whole relic extension, we can conclude that the observed polarization level can be reached only with the contribution of the polarized emission from the relic.
\par Another possible reason for the lack of polarized emission in the southern relic could be that the turbulence in the relic has mixed the magnetic field lines. Turbulence mixes field lines at scales larger than the Alfv\'{e}n  scale ($l_{A}$), the scale at which the velocity of turbulence is equal to Alfv\'{e}n speed~\citep{2016brunetti}. If this scale is smaller than the beam size, then we would observe depolarization effects. In view of this, the stark differences between the northern and southern relics may be explained if the turbulence in the north relic is generated with lower efficiency, or at larger scales than in the southern relic, or has been already partly dissipated.
\par In Section~\ref{subsec:mfsrmsynthesisresults}, we presented the results obtained with new observations and techniques. With our analysis, we found an increase of the polarization fraction for the southern relic ($\sim14.6\%$) with respect to what was found in previous work~\citep[values below the $10\%$,][]{2021jones}. This is still below the average reported for such systems~\citep[][$~\sim20\%$]{2022stuardi}, and also below the polarization of the northern relic. RM synthesis is able to solve for the bandwidth depolarization. Given the absence of systematic and significant pixels that show complex Faraday spectra (see Appendix~\ref{app2}), we infer that the depolarization of the southern relic has an external nature, related to the properties of the magnetized ICM that is crossed by the relic emission. Unfortunately, the complex morphology of the X-rays emission, together with the low counts statistics~\citep{2021finner, tumer2023}, does not allow to reconstruct the density distribution of the thermal gas within the cluster, that is crucial to quantify the effect of the Faraday rotation as well as to recover information on the magnetic field along the line of sight (Eq.~\ref{eq:faradaydepth}). Another important factor that can determine a difference in the polarization properties between the two relics is their three-dimensional position within the cluster. What we observe is the projection of the relics on the plane of the sky but they present a rather complex morphology~\citep{wittor2019}, so variations in RM can be addressed to the fact that different parts of the relic can be at different distances within the cluster, and so the radiation crosses different lengths of the ICM. Given the lack of detailed information currently available on the gas distribution and the inability to get geometric hints from observations we cannot distinguish clearly the contribution of the two scenarios just discussed to the relic depolarization. Based on the results of our RM study with respect to previous papers, we can discard the hypothesis~\citep[raised in][]{2014degasperin} that the southern relic lies further from us, at least for what concerns the portion of the relic observed in polarization. Nevertheless, we cannot exclude that the southern relic is a more complex system, with some parts that being located deeper through the ICM are completely depolarized at 13". From the resulting values of RM, similar between the two relics, and the depolarization trend observed for the southern relic, we favor a scenario in which the southern relic is morphologically more complex and/or located in a more turbulent region of the cluster.

\subsection{Constraints on the magnetic field of the cluster}

After discussing the origin of the depolarization for the southern relic, we used the beam depolarization to get constraints on the magnetic field distribution within the cluster. This strategy has been already adopted in other papers~\citep{2011bonafede, 2022osinga}. When synchrotron emission arising from a cluster or background source crosses the ICM, regions with similar intrinsic polarization angles, going through different paths, will be subject to differential Faraday rotation. If the magnetic field in the foreground screen is tangled on scales much smaller than the observing beam, then radiation with similar $\phi$ but opposite sign will be averaged out, and the observed degree of polarization will be reduced.
\par We reproduced the observed beam depolarization trend for the southern relic with simulations of the cluster magnetic power spectrum. This, as already stated previously, relies on several assumptions, mainly due to the fact that we do not know the true ICM distribution as well as the true relic structure, that directly affects the computation of the equipartition magnetic field.  The equipartition magnetic field computed for the southern relic has been used as a reference to set $B_{\rm{mean}}$ for the simulations. From the comparison between data and simulations shown in Fig.~\ref{fig:PSZ096_obs_vs_sims} we clearly see that the Kolmogorov power spectrum is not able to reproduce both the normalization of the highest resolution point and the fast depolarization before reaching the plateau at low resolutions. Moreover, the value of polarization fraction evaluated at $25"$ is not fitted by any model used for the magnetic field spectrum. Overall, the PDF power spectrum, based on cosmological simulations, is the one that best fits the observed beam depolarization for the S relic. Among the wide range of scales tested, we found that the best combination of minimum and maximum scale is $\Lambda_{\rm{min}} = 35~\rm{kpc}$ and $\Lambda_{\rm{max}} = 400~\rm{kpc}$, with a $B_{\rm{mean}}$ set to $0.8~\rm{\mu G}$: this returns in an average magnetic field over 1 Mpc of $\sim 2~\rm{\mu G}$. 

\section{Conclusions}

In this paper, we address the origin of the low fractional polarization that characterizes the southern relic of PSZ2 G096.88+24.18, a galaxy cluster hosting a double radio relic. While the northern relic shows a rather common behavior compared to other radio relics (both in terms of fractional polarization and orientation of the magnetic field lines) the southern relic shows a lack of polarization, as already reported in previous work~\citep{2014degasperin, 2021jones}, as well as an opposite orientation of the magnetic field vectors with respect to what expected in presence of a shock~\citep{2019vanweeren}. We used new and archival VLA observations in L-band using multiple configurations, in order to achieve both high angular resolution and sensitivity to the extended emission. We analyzed the integrated polarization fraction of both relics and we confirmed the low fractional polarization that characterizes the southern relic. To correct for bandwidth depolarization, we used the rotation measure synthesis technique for the polarization analysis. With this technique, we found higher values of fractional polarization for both relics (18.6\% for the northern relic and 14.6\% for the southern relic), yet the south relic shows lower fractional polarization with respect to the northern one. The two relics show different depolarization trends with increasing beam size, with the southern relic being the one with the greater decrease in polarization fraction. With this RM study, we also inferred a higher average value of RM for the southern relic with respect to the north, which however has a higher $\sigma_{\rm{RM}}$: this means that the RM is more uniform in the southern relic. This result depends also on the fact that we detected RM only in a few pixels, and mostly for the northern relic.
\par We infer that the depolarization has an external nature, related to the different composition or different distribution of the Faraday screen, other that the true structure and position of the relics within the cluster. Unfortunately, given the complexity of the X-ray emission from the ICM, its density distribution has not been determined yet, so we do not have a direct estimate for $n_{e}$, that is crucial to recover information on the magnetic field along the line-of-sight (Eq.~\ref{eq:faradaydepth}).
\par We also constrained the magnetic field power spectrum within the cluster using the beam depolarization observed for the S relic. Computing the fractional polarization trend with respect to the beam size with different models, we found that the best combination of parameters that could fit the observations is given by the magnetic field power spectrum presented in~\citet{2019dominguezfernandez} defined on scales $\Lambda_{\rm{min}} = 35~\rm{kpc}$ and $\Lambda_{\rm{max}} = 400~\rm{kpc}$. This yields a magnetic field averaged over 1 Mpc scale of $\sim 2~\rm{\mu G}$, that is in line to what is typically found in other clusters, using the Faraday rotation effect for similar studies~\citep{2004murgia, 2010bonafede, 2022stuardi}.

\begin{acknowledgements}
We would like to thank our anonymous referee for the useful comments on this manuscript. We thank A. Botteon for sharing the X-ray data shown in Fig.~\ref{fig:PSZ096_totalintensity_10asec}. This paper is supported by the Fondazione ICSC, Spoke 3 Astrophysics and Cosmos Observations. National Recovery and Resilience Plan (Piano Nazionale di Ripresa e Resilienza, PNRR) Project ID CN\_00000013 "Italian Research Center for High-Performance Computing, Big Data and Quantum Computing" funded by MUR Missione 4 Componente 2 Investimento 1.4: Potenziamento strutture di ricerca e creazione di "campioni nazionali di R\&S (M4C2-19)" - Next Generation EU (NGEU). AB acknowledges support from the ERC-Stg DRANOEL n. 714245. FV has been supported by Fondazione Cariplo and Fondazione CDP, through grant n. Rif: 2022-2088 CUP J33C22004310003 for "BREAKTHRU" project. FdG acknowledges support from the ERC Consolidator Grant ULU 101086378. RJvW acknowledges support from the ERC Starting Grant ClusterWeb 804208. The National Radio Astronomy Observatory is a facility of the National Science Foundation operated under cooperative agreement by Associated Universities, Inc.
\end{acknowledgements}


\bibliographystyle{aa}
\bibliography{example}

\begin{thebibliography}{56}
\expandafter\ifx\csname natexlab\endcsname\relax\def\natexlab#1{#1}\fi

\bibitem[{{Akamatsu} \& {Kawahara}(2013)}]{akamatsu2013}
{Akamatsu}, H. \& {Kawahara}, H. 2013, \pasj, 65, 16

\bibitem[{{B{\"o}hringer} {et~al.}(2016){B{\"o}hringer}, {Chon}, \&
  {Kronberg}}]{2016bohringer}
{B{\"o}hringer}, H., {Chon}, G., \& {Kronberg}, P.~P. 2016, Astronomy \&
  Astrophysics, 596, A22

\bibitem[{{Bonafede} {et~al.}(2012){Bonafede}, {Br{\"u}ggen}, {van Weeren},
  {Vazza}, {Giovannini}, {Ebeling}, {Edge}, {Hoeft}, \& {Klein}}]{2012bonafede}
{Bonafede}, A., {Br{\"u}ggen}, M., {van Weeren}, R., {et~al.} 2012, Monthly
  Notices of the Royal Astronomical Society, 426, 40

\bibitem[{{Bonafede} {et~al.}(2010){Bonafede}, {Feretti}, {Murgia}, {Govoni},
  {Giovannini}, {Dallacasa}, {Dolag}, \& {Taylor}}]{2010bonafede}
{Bonafede}, A., {Feretti}, L., {Murgia}, M., {et~al.} 2010, Astronomy and
  Astrophysics, 513, A30

\bibitem[{{Bonafede} {et~al.}(2011){Bonafede}, {Govoni}, {Feretti}, {Murgia},
  {Giovannini}, \& {Br{\"u}ggen}}]{2011bonafede}
{Bonafede}, A., {Govoni}, F., {Feretti}, L., {et~al.} 2011, Astronomy and
  Astrophysics, 530, A24

\bibitem[{{Bonafede} {et~al.}(2013){Bonafede}, {Vazza}, {Br{\"u}ggen},
  {Murgia}, {Govoni}, {Feretti}, {Giovannini}, \& {Ogrean}}]{2013bonafede}
{Bonafede}, A., {Vazza}, F., {Br{\"u}ggen}, M., {et~al.} 2013, Monthly Notices
  of the Royal Astronomical Society, 433, 3208

\bibitem[{{Botteon} {et~al.}(2020){Botteon}, {Brunetti}, {Ryu}, \&
  {Roh}}]{botteon2020}
{Botteon}, A., {Brunetti}, G., {Ryu}, D., \& {Roh}, S. 2020, \aap, 634, A64

\bibitem[{{Brentjens} \& {de Bruyn}(2005)}]{2005brentjens}
{Brentjens}, M.~A. \& {de Bruyn}, A.~G. 2005, Astronomy and Astrophysics, 441,
  1217

\bibitem[{{Briggs}(1995)}]{1995briggs}
{Briggs}, D.~S. 1995, PhD thesis, New Mexico Institute of Mining and Technology

\bibitem[{{Br{\"u}ggen}(2013)}]{2013bruggen}
{Br{\"u}ggen}, M. 2013, Astronomische Nachrichten, 334, 543

\bibitem[{{Brunetti} \& {Jones}(2014)}]{2014brunetti}
{Brunetti}, G. \& {Jones}, T.~W. 2014, International Journal of Modern Physics
  D, 23, 1430007

\bibitem[{{Brunetti} \& {Lazarian}(2016)}]{2016brunetti}
{Brunetti}, G. \& {Lazarian}, A. 2016, Monthly Notices of the Royal
  Astronomical Society, 458, 2584

\bibitem[{{Burn}(1966)}]{1966burn}
{Burn}, B.~J. 1966, Monthly Notices of the Royal Astronomical Society, 133, 67

\bibitem[{{Cornwell} {et~al.}(2008){Cornwell}, {Golap}, \&
  {Bhatnagar}}]{2008cornwell}
{Cornwell}, T.~J., {Golap}, K., \& {Bhatnagar}, S. 2008, IEEE Journal of
  Selected Topics in Signal Processing, 2, 647

\bibitem[{{de Gasperin} {et~al.}(2022){de Gasperin}, {Rudnick}, {Finoguenov},
  {Wittor}, {Akamatsu}, {Br{\"u}ggen}, {Chibueze}, {Clarke}, {Cotton},
  {Cuciti}, {Dom{\'\i}nguez-Fern{\'a}ndez}, {Knowles}, {O'Sullivan}, \&
  {Sebokolodi}}]{2022degasperin}
{de Gasperin}, F., {Rudnick}, L., {Finoguenov}, A., {et~al.} 2022, Astronomy
  and Astrophysics, 659, A146

\bibitem[{{de Gasperin} {et~al.}(2014){de Gasperin}, {van Weeren},
  {Br{\"u}ggen}, {Vazza}, {Bonafede}, \& {Intema}}]{2014degasperin}
{de Gasperin}, F., {van Weeren}, R.~J., {Br{\"u}ggen}, M., {et~al.} 2014,
  Monthly Notices of the Royal Astronomical Society, 444, 3130

\bibitem[{{Di Gennaro} {et~al.}(2021){Di Gennaro}, {van Weeren}, {Rudnick},
  {Hoeft}, {Br{\"u}ggen}, {Ryu}, {R{\"o}ttgering}, {Forman}, {Stroe},
  {Shimwell}, {Kraft}, {Jones}, \& {Hoang}}]{2021digennaro}
{Di Gennaro}, G., {van Weeren}, R.~J., {Rudnick}, L., {et~al.} 2021,
  Astrophysical Journal, 911, 3

\bibitem[{{Dom{\'\i}nguez-Fern{\'a}ndez}
  {et~al.}(2019){Dom{\'\i}nguez-Fern{\'a}ndez}, {Vazza}, {Br{\"u}ggen}, \&
  {Brunetti}}]{2019dominguezfernandez}
{Dom{\'\i}nguez-Fern{\'a}ndez}, P., {Vazza}, F., {Br{\"u}ggen}, M., \&
  {Brunetti}, G. 2019, Monthly Notices of the Royal Astronomical Society, 486,
  623

\bibitem[{{En{\ss}lin} {et~al.}(1998){En{\ss}lin}, {Biermann}, {Klein}, \&
  {Kohle}}]{1998enblin}
{En{\ss}lin}, T.~A., {Biermann}, P.~L., {Klein}, U., \& {Kohle}, S. 1998,
  Astronomy and Astrophysics, 332, 395

\bibitem[{{Feretti} {et~al.}(2012){Feretti}, {Giovannini}, {Govoni}, \&
  {Murgia}}]{feretti2012}
{Feretti}, L., {Giovannini}, G., {Govoni}, F., \& {Murgia}, M. 2012, \aapr, 20,
  54

\bibitem[{{Finner} {et~al.}(2021){Finner}, {HyeongHan}, {Jee}, {Wittman},
  {Forman}, {van Weeren}, {Golovich}, {Dawson}, {Jones}, {de Gasperin}, \&
  {Jones}}]{2021finner}
{Finner}, K., {HyeongHan}, K., {Jee}, M.~J., {et~al.} 2021, Astrophysical
  Journal, 918, 72

\bibitem[{{George} {et~al.}(2012){George}, {Stil}, \& {Keller}}]{2012george}
{George}, S.~J., {Stil}, J.~M., \& {Keller}, B.~W. 2012, Publications of the
  Astronomical Society of Australia, 29, 214

\bibitem[{{Golovich} {et~al.}(2019){Golovich}, {Dawson}, {Wittman}, {Jee},
  {Benson}, {Lemaux}, {van Weeren}, {Andrade-Santos}, {Sobral}, {de Gasperin},
  {Br{\"u}ggen}, {Brada{\v{c}}}, {Finner}, \& {Peter}}]{2019golovich}
{Golovich}, N., {Dawson}, W.~A., {Wittman}, D.~M., {et~al.} 2019, Astrophysical
  Journal, Supplement, 240, 39

\bibitem[{{Govoni} \& {Feretti}(2004)}]{2004govoni}
{Govoni}, F. \& {Feretti}, L. 2004, International Journal of Modern Physics D,
  13, 1549

\bibitem[{{Govoni} {et~al.}(2006){Govoni}, {Murgia}, {Feretti}, {Giovannini},
  {Dolag}, \& {Taylor}}]{govoni2006}
{Govoni}, F., {Murgia}, M., {Feretti}, L., {et~al.} 2006, Astronomy and
  Astrophysics, 460, 425

\bibitem[{{Hales} {et~al.}(2012){Hales}, {Gaensler}, {Norris}, \&
  {Middelberg}}]{2012hales}
{Hales}, C.~A., {Gaensler}, B.~M., {Norris}, R.~P., \& {Middelberg}, E. 2012,
  Monthly Notices of the Royal Astronomical Society, 424, 2160

\bibitem[{{Hoang} {et~al.}(2017){Hoang}, {Shimwell}, {Stroe}, {Akamatsu},
  {Brunetti}, {Donnert}, {Intema}, {Mulcahy}, {R{\"o}ttgering}, {van Weeren},
  {Bonafede}, {Br{\"u}ggen}, {Cassano}, {Chy{\.z}y}, {En{\ss}lin}, {Ferrari},
  {de Gasperin}, {Gu}, {Hoeft}, {Miley}, {Orr{\'u}}, {Pizzo}, \&
  {White}}]{2017hoang}
{Hoang}, D.~N., {Shimwell}, T.~W., {Stroe}, A., {et~al.} 2017, Monthly Notices
  of the Royal Astronomical Society, 471, 1107

\bibitem[{{Jones} {et~al.}(2021){Jones}, {de Gasperin}, {Cuciti}, {Hoang},
  {Botteon}, {Br{\"u}ggen}, {Brunetti}, {Finner}, {Forman}, {Jones}, {Kraft},
  {Shimwell}, \& {van Weeren}}]{2021jones}
{Jones}, A., {de Gasperin}, F., {Cuciti}, V., {et~al.} 2021, Monthly Notices of
  the Royal Astronomical Society

\bibitem[{{McMullin} {et~al.}(2007){McMullin}, {Waters}, {Schiebel}, {Young},
  \& {Golap}}]{2007casa}
{McMullin}, J.~P., {Waters}, B., {Schiebel}, D., {Young}, W., \& {Golap}, K.
  2007, in Astronomical Society of the Pacific Conference Series, Vol. 376,
  Astronomical Data Analysis Software and Systems XVI, ed. R.~A. {Shaw},
  F.~{Hill}, \& D.~J. {Bell}, 127

\bibitem[{{Murgia} {et~al.}(2004){Murgia}, {Govoni}, {Feretti}, {Giovannini},
  {Dallacasa}, {Fanti}, {Taylor}, \& {Dolag}}]{2004murgia}
{Murgia}, M., {Govoni}, F., {Feretti}, L., {et~al.} 2004, Astronomy and
  Astrophysics, 424, 429

\bibitem[{{Offringa} {et~al.}(2014){Offringa}, {McKinley}, {Hurley-Walker},
  {Briggs}, {Wayth}, {Kaplan}, {Bell}, {Feng}, {Neben}, {Hughes}, {Rhee},
  {Murphy}, {Bhat}, {Bernardi}, {Bowman}, {Cappallo}, {Corey}, {Deshpande},
  {Emrich}, {Ewall-Wice}, {Gaensler}, {Goeke}, {Greenhill}, {Hazelton},
  {Hindson}, {Johnston-Hollitt}, {Jacobs}, {Kasper}, {Kratzenberg}, {Lenc},
  {Lonsdale}, {Lynch}, {McWhirter}, {Mitchell}, {Morales}, {Morgan},
  {Kudryavtseva}, {Oberoi}, {Ord}, {Pindor}, {Procopio}, {Prabu}, {Riding},
  {Roshi}, {Shankar}, {Srivani}, {Subrahmanyan}, {Tingay}, {Waterson},
  {Webster}, {Whitney}, {Williams}, \& {Williams}}]{2014offringa}
{Offringa}, A.~R., {McKinley}, B., {Hurley-Walker}, N., {et~al.} 2014, Monthly
  Notices of the Royal Astronomical Society, 444, 606

\bibitem[{{Oppermann} {et~al.}(2015){Oppermann}, {Junklewitz}, {Greiner},
  {En{\ss}lin}, {Akahori}, {Carretti}, {Gaensler}, {Goobar}, {Harvey-Smith},
  {Johnston-Hollitt}, {Pratley}, {Schnitzeler}, {Stil}, \&
  {Vacca}}]{oppermann2015}
{Oppermann}, N., {Junklewitz}, H., {Greiner}, M., {et~al.} 2015, \aap, 575,
  A118

\bibitem[{{Osinga} {et~al.}(2022){Osinga}, {van Weeren}, {Andrade-Santos},
  {Rudnick}, {Bonafede}, {Clarke}, {Duncan}, {Giacintucci}, {Mroczkowski}, \&
  {R{\"o}ttgering}}]{2022osinga}
{Osinga}, E., {van Weeren}, R.~J., {Andrade-Santos}, F., {et~al.} 2022, arXiv
  e-prints, arXiv:2207.09717

\bibitem[{{O'Sullivan} {et~al.}(2012){O'Sullivan}, {Brown}, {Robishaw},
  {Schnitzeler}, {McClure-Griffiths}, {Feain}, {Taylor}, {Gaensler},
  {Landecker}, {Harvey-Smith}, \& {Carretti}}]{2012osullivan}
{O'Sullivan}, S.~P., {Brown}, S., {Robishaw}, T., {et~al.} 2012, Monthly
  Notices of the Royal Astronomical Society, 421, 3300

\bibitem[{{Perley} \& {Butler}(2013)}]{Perley13}
{Perley}, R.~A. \& {Butler}, B.~J. 2013, The Astrophysical Journal Supplement
  Series, 206, 16

\bibitem[{{Planck Collaboration} {et~al.}(2011){Planck Collaboration}, {Ade},
  {Aghanim}, {Arnaud}, {Ashdown}, {Aumont}, {Baccigalupi}, {Balbi}, {Banday},
  {Barreiro}, {Bartelmann}, {Bartlett}, {Battaner}, {Battye}, {Benabed},
  {Beno{\^\i}t}, {Bernard}, {Bersanelli}, {Bhatia}, {Bock}, {Bonaldi}, {Bond},
  {Borrill}, {Bouchet}, {Brown}, {Bucher}, {Burigana}, {Cabella}, {Cantalupo},
  {Cardoso}, {Carvalho}, {Catalano}, {Cay{\'o}n}, {Challinor}, {Chamballu},
  {Chary}, {Chiang}, {Chiang}, {Chon}, {Christensen}, {Churazov}, {Clements},
  {Colafrancesco}, {Colombi}, {Couchot}, {Coulais}, {Crill}, {Cuttaia}, {da
  Silva}, {Dahle}, {Danese}, {Davis}, {de Bernardis}, {de Gasperis}, {de Rosa},
  {de Zotti}, {Delabrouille}, {Delouis}, {D{\'e}sert}, {Dickinson}, {Diego},
  {Dolag}, {Dole}, {Donzelli}, {Dor{\'e}}, {D{\"o}rl}, {Douspis}, {Dupac},
  {Efstathiou}, {Eisenhardt}, {En{\ss}lin}, {Feroz}, {Finelli}, {Flores-Cacho},
  {Forni}, {Fosalba}, {Frailis}, {Franceschi}, {Fromenteau}, {Galeotta},
  {Ganga}, {G{\'e}nova-Santos}, {Giard}, {Giardino}, {Giraud-H{\'e}raud},
  {Gonz{\'a}lez-Nuevo}, {Gonz{\'a}lez-Riestra}, {G{\'o}rski}, {Grainge},
  {Gratton}, {Gregorio}, {Gruppuso}, {Harrison}, {Hein{\"a}m{\"a}ki},
  {Henrot-Versill{\'e}}, {Hern{\'a}ndez-Monteagudo}, {Herranz}, {Hildebrandt},
  {Hivon}, {Hobson}, {Holmes}, {Hovest}, {Hoyland}, {Huffenberger}, {Hurier},
  {Hurley-Walker}, {Jaffe}, {Jones}, {Juvela}, {Keih{\"a}nen}, {Keskitalo},
  {Kisner}, {Kneissl}, {Knox}, {Kurki-Suonio}, {Lagache}, {Lamarre}, {Lasenby},
  {Laureijs}, {Lawrence}, {Le Jeune}, {Leach}, {Leonardi}, {Li}, {Liddle},
  {Lilje}, {Linden-V{\o}rnle}, {L{\'o}pez-Caniego}, {Lubin},
  {Mac{\'\i}as-P{\'e}rez}, {MacTavish}, {Maffei}, {Maino}, {Mandolesi}, {Mann},
  {Maris}, {Marleau}, {Mart{\'\i}nez-Gonz{\'a}lez}, {Masi}, {Matarrese},
  {Matthai}, {Mazzotta}, {Mei}, {Meinhold}, {Melchiorri}, {Melin}, {Mendes},
  {Mennella}, {Mitra}, {Miville-Desch{\^e}nes}, {Moneti}, {Montier},
  {Morgante}, {Mortlock}, {Munshi}, {Murphy}, {Naselsky}, {Nati}, {Natoli},
  {Netterfield}, {N{\o}rgaard-Nielsen}, {Noviello}, {Novikov}, {Novikov},
  {Olamaie}, {Osborne}, {Pajot}, {Pasian}, {Patanchon}, {Pearson}, {Perdereau},
  {Perotto}, {Perrotta}, {Piacentini}, {Piat}, {Pierpaoli}, {Piffaretti},
  {Plaszczynski}, {Pointecouteau}, {Polenta}, {Ponthieu}, {Poutanen}, {Pratt},
  {Pr{\'e}zeau}, {Prunet}, {Puget}, {Rachen}, {Reach}, {Rebolo}, {Reinecke},
  {Renault}, {Ricciardi}, {Riller}, {Ristorcelli}, {Rocha}, {Rosset},
  {Rubi{\~n}o-Mart{\'\i}n}, {Rusholme}, {Saar}, {Sandri}, {Santos}, {Saunders},
  {Savini}, {Schaefer}, {Scott}, {Seiffert}, {Shellard}, {Smoot}, {Stanford},
  {Starck}, {Stivoli}, {Stolyarov}, {Stompor}, {Sudiwala}, {Sunyaev}, {Sutton},
  {Sygnet}, {Taburet}, {Tauber}, {Terenzi}, {Toffolatti}, {Tomasi}, {Torre},
  {Tristram}, {Tuovinen}, {Valenziano}, {Vibert}, {Vielva}, {Villa},
  {Vittorio}, {Wade}, {Wandelt}, {Weller}, {White}, {White}, {Yvon}, {Zacchei},
  \& {Zonca}}]{2011pszcatalogue}
{Planck Collaboration}, {Ade}, P.~A.~R., {Aghanim}, N., {et~al.} 2011,
  Astronomy and Astrophysics, 536, A8

\bibitem[{{Planck Collaboration} {et~al.}(2016){Planck Collaboration}, {Ade},
  {Aghanim}, {Arnaud}, {Ashdown}, {Aumont}, {Baccigalupi}, {Banday},
  {Barreiro}, {Barrena}, {Bartlett}, {Bartolo}, {Battaner}, {Battye},
  {Benabed}, {Beno{\^\i}t}, {Benoit-L{\'e}vy}, {Bernard}, {Bersanelli},
  {Bielewicz}, {Bikmaev}, {B{\"o}hringer}, {Bonaldi}, {Bonavera}, {Bond},
  {Borrill}, {Bouchet}, {Bucher}, {Burenin}, {Burigana}, {Butler}, {Calabrese},
  {Cardoso}, {Carvalho}, {Catalano}, {Challinor}, {Chamballu}, {Chary},
  {Chiang}, {Chon}, {Christensen}, {Clements}, {Colombi}, {Colombo}, {Combet},
  {Comis}, {Couchot}, {Coulais}, {Crill}, {Curto}, {Cuttaia}, {Dahle},
  {Danese}, {Davies}, {Davis}, {de Bernardis}, {de Rosa}, {de Zotti},
  {Delabrouille}, {D{\'e}sert}, {Dickinson}, {Diego}, {Dolag}, {Dole},
  {Donzelli}, {Dor{\'e}}, {Douspis}, {Ducout}, {Dupac}, {Efstathiou},
  {Eisenhardt}, {Elsner}, {En{\ss}lin}, {Eriksen}, {Falgarone}, {Fergusson},
  {Feroz}, {Ferragamo}, {Finelli}, {Forni}, {Frailis}, {Fraisse}, {Franceschi},
  {Frejsel}, {Galeotta}, {Galli}, {Ganga}, {G{\'e}nova-Santos}, {Giard},
  {Giraud-H{\'e}raud}, {Gjerl{\o}w}, {Gonz{\'a}lez-Nuevo}, {G{\'o}rski},
  {Grainge}, {Gratton}, {Gregorio}, {Gruppuso}, {Gudmundsson}, {Hansen},
  {Hanson}, {Harrison}, {Hempel}, {Henrot-Versill{\'e}},
  {Hern{\'a}ndez-Monteagudo}, {Herranz}, {Hildebrandt}, {Hivon}, {Hobson},
  {Holmes}, {Hornstrup}, {Hovest}, {Huffenberger}, {Hurier}, {Jaffe}, {Jaffe},
  {Jin}, {Jones}, {Juvela}, {Keih{\"a}nen}, {Keskitalo}, {Khamitov}, {Kisner},
  {Kneissl}, {Knoche}, {Kunz}, {Kurki-Suonio}, {Lagache}, {Lamarre}, {Lasenby},
  {Lattanzi}, {Lawrence}, {Leonardi}, {Lesgourgues}, {Levrier}, {Liguori},
  {Lilje}, {Linden-V{\o}rnle}, {L{\'o}pez-Caniego}, {Lubin},
  {Mac{\'\i}as-P{\'e}rez}, {Maggio}, {Maino}, {Mak}, {Mandolesi}, {Mangilli},
  {Martin}, {Mart{\'\i}nez-Gonz{\'a}lez}, {Masi}, {Matarrese}, {Mazzotta},
  {McGehee}, {Mei}, {Melchiorri}, {Melin}, {Mendes}, {Mennella}, {Migliaccio},
  {Mitra}, {Miville-Desch{\^e}nes}, {Moneti}, {Montier}, {Morgante},
  {Mortlock}, {Moss}, {Munshi}, {Murphy}, {Naselsky}, {Nastasi}, {Nati},
  {Natoli}, {Netterfield}, {N{\o}rgaard-Nielsen}, {Noviello}, {Novikov},
  {Novikov}, {Olamaie}, {Oxborrow}, {Paci}, {Pagano}, {Pajot}, {Paoletti},
  {Pasian}, {Patanchon}, {Pearson}, {Perdereau}, {Perotto}, {Perrott},
  {Perrotta}, {Pettorino}, {Piacentini}, {Piat}, {Pierpaoli}, {Pietrobon},
  {Plaszczynski}, {Pointecouteau}, {Polenta}, {Pratt}, {Pr{\'e}zeau}, {Prunet},
  {Puget}, {Rachen}, {Reach}, {Rebolo}, {Reinecke}, {Remazeilles}, {Renault},
  {Renzi}, {Ristorcelli}, {Rocha}, {Rosset}, {Rossetti}, {Roudier}, {Rozo},
  {Rubi{\~n}o-Mart{\'\i}n}, {Rumsey}, {Rusholme}, {Rykoff}, {Sandri}, {Santos},
  {Saunders}, {Savelainen}, {Savini}, {Schammel}, {Scott}, {Seiffert},
  {Shellard}, {Shimwell}, {Spencer}, {Stanford}, {Stern}, {Stolyarov},
  {Stompor}, {Streblyanska}, {Sudiwala}, {Sunyaev}, {Sutton}, {Suur-Uski},
  {Sygnet}, {Tauber}, {Terenzi}, {Toffolatti}, {Tomasi}, {Tramonte},
  {Tristram}, {Tucci}, {Tuovinen}, {Umana}, {Valenziano}, {Valiviita}, {Van
  Tent}, {Vielva}, {Villa}, {Wade}, {Wandelt}, {Wehus}, {White}, {Wright},
  {Yvon}, {Zacchei}, \& {Zonca}}]{2016psz2catalogue}
{Planck Collaboration}, {Ade}, P.~A.~R., {Aghanim}, N., {et~al.} 2016,
  Astronomy and Astrophysics, 594, A27

\bibitem[{{Purcell} {et~al.}(2020){Purcell}, {Van Eck}, {West}, {Sun}, \&
  {Gaensler}}]{2020rmtools}
{Purcell}, C.~R., {Van Eck}, C.~L., {West}, J., {Sun}, X.~H., \& {Gaensler},
  B.~M. 2020, {RM-Tools: Rotation measure (RM) synthesis and Stokes
  QU-fitting}, Astrophysics Source Code Library, record ascl:2005.003

\bibitem[{{Rajpurohit} {et~al.}(2022){Rajpurohit}, {van Weeren}, {Hoeft},
  {Vazza}, {Brienza}, {Forman}, {Wittor}, {Dom{\'\i}nguez-Fern{\'a}ndez},
  {Rajpurohit}, {Riseley}, {Botteon}, {Osinga}, {Brunetti}, {Bonnassieux},
  {Bonafede}, {Rajpurohit}, {Stuardi}, {Drabent}, {Br{\"u}ggen}, {Dallacasa},
  {Shimwell}, {R{\"o}ttgering}, {Gasperin}, {Miley}, \&
  {Rossetti}}]{2022rajpurohit}
{Rajpurohit}, K., {van Weeren}, R.~J., {Hoeft}, M., {et~al.} 2022,
  Astrophysical Journal, 927, 80

\bibitem[{{Rau} \& {Cornwell}(2011)}]{2011rau}
{Rau}, U. \& {Cornwell}, T.~J. 2011, Astronomy and Astrophysics, 532, A71

\bibitem[{{Rincon}(2019)}]{2019JPlPh..85d2001R}
{Rincon}, F. 2019, Journal of Plasma Physics, 85, 205850401

\bibitem[{{Sokoloff} {et~al.}(1998){Sokoloff}, {Bykov}, {Shukurov},
  {Berkhuijsen}, {Beck}, \& {Poezd}}]{1998sokoloff}
{Sokoloff}, D.~D., {Bykov}, A.~A., {Shukurov}, A., {et~al.} 1998, Monthly
  Notices of the Royal Astronomical Society, 299, 189

\bibitem[{{Stuardi} {et~al.}(2021){Stuardi}, {Bonafede}, {Lovisari},
  {Dom{\'\i}nguez-Fern{\'a}ndez}, {Vazza}, {Br{\"u}ggen}, {van Weeren}, \& {de
  Gasperin}}]{2021stuardi}
{Stuardi}, C., {Bonafede}, A., {Lovisari}, L., {et~al.} 2021, Monthly Notices
  of the Royal Astronomical Society, 502, 2518

\bibitem[{{Stuardi} {et~al.}(2022){Stuardi}, {Bonafede}, {Rajpurohit},
  {Br{\"u}ggen}, {de Gasperin}, {Hoang}, {van Weeren}, \&
  {Vazza}}]{2022stuardi}
{Stuardi}, C., {Bonafede}, A., {Rajpurohit}, K., {et~al.} 2022, arXiv e-prints,
  arXiv:2207.00503

\bibitem[{{Stuardi} {et~al.}(2019){Stuardi}, {Bonafede}, {Wittor}, {Vazza},
  {Botteon}, {Locatelli}, {Dallacasa}, {Golovich}, {Hoeft}, {van Weeren},
  {Br{\"u}ggen}, \& {de Gasperin}}]{stuardi2019}
{Stuardi}, C., {Bonafede}, A., {Wittor}, D., {et~al.} 2019, \mnras, 489, 3905

\bibitem[{{Tribble}(1991)}]{1991tribble}
{Tribble}, P.~C. 1991, Monthly Notices of the Royal Astronomical Society, 253,
  147

\bibitem[{{T{\"u}mer} {et~al.}(2023){T{\"u}mer}, {Wik}, {Schellenberger},
  {Miller}, \& {Bautz}}]{tumer2023}
{T{\"u}mer}, A., {Wik}, D.~R., {Schellenberger}, G., {Miller}, E.~D., \&
  {Bautz}, M.~W. 2023, arXiv e-prints, arXiv:2312.06020

\bibitem[{{Vacca} {et~al.}(2018){Vacca}, {Murgia}, {Govoni}, {En{\ss}lin},
  {Oppermann}, {Feretti}, {Giovannini}, \& {Loi}}]{2018vacca}
{Vacca}, V., {Murgia}, M., {Govoni}, F., {et~al.} 2018, Galaxies, 6, 142

\bibitem[{{van Haarlem} {et~al.}(2013){van Haarlem}, {Wise}, {Gunst}, {Heald},
  {McKean}, {Hessels}, {de Bruyn}, {Nijboer}, {Swinbank}, {Fallows},
  {Brentjens}, {Nelles}, {Beck}, {Falcke}, {Fender}, {H{\"o}randel},
  {Koopmans}, {Mann}, {Miley}, {R{\"o}ttgering}, {Stappers}, {Wijers},
  {Zaroubi}, {van den Akker}, {Alexov}, {Anderson}, {Anderson}, {van Ardenne},
  {Arts}, {Asgekar}, {Avruch}, {Batejat}, {B{\"a}hren}, {Bell}, {Bell}, {van
  Bemmel}, {Bennema}, {Bentum}, {Bernardi}, {Best}, {B{\^\i}rzan}, {Bonafede},
  {Boonstra}, {Braun}, {Bregman}, {Breitling}, {van de Brink}, {Broderick},
  {Broekema}, {Brouw}, {Br{\"u}ggen}, {Butcher}, {van Cappellen}, {Ciardi},
  {Coenen}, {Conway}, {Coolen}, {Corstanje}, {Damstra}, {Davies}, {Deller},
  {Dettmar}, {van Diepen}, {Dijkstra}, {Donker}, {Doorduin}, {Dromer}, {Drost},
  {van Duin}, {Eisl{\"o}ffel}, {van Enst}, {Ferrari}, {Frieswijk}, {Gankema},
  {Garrett}, {de Gasperin}, {Gerbers}, {de Geus}, {Grie{\ss}meier}, {Grit},
  {Gruppen}, {Hamaker}, {Hassall}, {Hoeft}, {Holties}, {Horneffer}, {van der
  Horst}, {van Houwelingen}, {Huijgen}, {Iacobelli}, {Intema}, {Jackson},
  {Jelic}, {de Jong}, {Juette}, {Kant}, {Karastergiou}, {Koers}, {Kollen},
  {Kondratiev}, {Kooistra}, {Koopman}, {Koster}, {Kuniyoshi}, {Kramer},
  {Kuper}, {Lambropoulos}, {Law}, {van Leeuwen}, {Lemaitre}, {Loose}, {Maat},
  {Macario}, {Markoff}, {Masters}, {McFadden}, {McKay-Bukowski}, {Meijering},
  {Meulman}, {Mevius}, {Middelberg}, {Millenaar}, {Miller-Jones}, {Mohan},
  {Mol}, {Morawietz}, {Morganti}, {Mulcahy}, {Mulder}, {Munk}, {Nieuwenhuis},
  {van Nieuwpoort}, {Noordam}, {Norden}, {Noutsos}, {Offringa}, {Olofsson},
  {Omar}, {Orr{\'u}}, {Overeem}, {Paas}, {Pandey-Pommier}, {Pandey}, {Pizzo},
  {Polatidis}, {Rafferty}, {Rawlings}, {Reich}, {de Reijer}, {Reitsma},
  {Renting}, {Riemers}, {Rol}, {Romein}, {Roosjen}, {Ruiter}, {Scaife}, {van
  der Schaaf}, {Scheers}, {Schellart}, {Schoenmakers}, {Schoonderbeek},
  {Serylak}, {Shulevski}, {Sluman}, {Smirnov}, {Sobey}, {Spreeuw}, {Steinmetz},
  {Sterks}, {Stiepel}, {Stuurwold}, {Tagger}, {Tang}, {Tasse}, {Thomas},
  {Thoudam}, {Toribio}, {van der Tol}, {Usov}, {van Veelen}, {van der Veen},
  {ter Veen}, {Verbiest}, {Vermeulen}, {Vermaas}, {Vocks}, {Vogt}, {de Vos},
  {van der Wal}, {van Weeren}, {Weggemans}, {Weltevrede}, {White}, {Wijnholds},
  {Wilhelmsson}, {Wucknitz}, {Yatawatta}, {Zarka}, {Zensus}, \& {van
  Zwieten}}]{vanhaarlem2013}
{van Haarlem}, M.~P., {Wise}, M.~W., {Gunst}, A.~W., {et~al.} 2013, \aap, 556,
  A2

\bibitem[{{van Weeren} {et~al.}(2016){van Weeren}, {Brunetti}, {Br{\"u}ggen},
  {Andrade-Santos}, {Ogrean}, {Williams}, {R{\"o}ttgering}, {Dawson}, {Forman},
  {de Gasperin}, {Hardcastle}, {Jones}, {Miley}, {Rafferty}, {Rudnick},
  {Sabater}, {Sarazin}, {Shimwell}, {Bonafede}, {Best}, {B{\^\i}rzan},
  {Cassano}, {Chy{\.z}y}, {Croston}, {Dijkema}, {En{\ss}lin}, {Ferrari},
  {Heald}, {Hoeft}, {Horellou}, {Jarvis}, {Kraft}, {Mevius}, {Intema},
  {Murray}, {Orr{\'u}}, {Pizzo}, {Sridhar}, {Simionescu}, {Stroe}, {van der
  Tol}, \& {White}}]{2016vanweeren}
{van Weeren}, R.~J., {Brunetti}, G., {Br{\"u}ggen}, M., {et~al.} 2016,
  Astrophysical Journal, 818, 204

\bibitem[{{van Weeren} {et~al.}(2019){van Weeren}, {de Gasperin}, {Akamatsu},
  {Br{\"u}ggen}, {Feretti}, {Kang}, {Stroe}, \& {Zandanel}}]{2019vanweeren}
{van Weeren}, R.~J., {de Gasperin}, F., {Akamatsu}, H., {et~al.} 2019, Space
  Science Reviews

\bibitem[{{van Weeren} {et~al.}(2010){van Weeren}, {R{\"o}ttgering},
  {Br{\"u}ggen}, \& {Hoeft}}]{2010vanweeren}
{van Weeren}, R.~J., {R{\"o}ttgering}, H. J.~A., {Br{\"u}ggen}, M., \& {Hoeft},
  M. 2010, Science, 330, 347

\bibitem[{{Vazza} {et~al.}(2018){Vazza}, {Brunetti}, {Br{\"u}ggen}, \&
  {Bonafede}}]{2018vazza}
{Vazza}, F., {Brunetti}, G., {Br{\"u}ggen}, M., \& {Bonafede}, A. 2018, Monthly
  Notices of the Royal Astronomical Society, 474, 1672

\bibitem[{{Wittor} {et~al.}(2019){Wittor}, {Hoeft}, {Vazza}, {Br{\"u}ggen}, \&
  {Dom{\'\i}nguez-Fern{\'a}ndez}}]{wittor2019}
{Wittor}, D., {Hoeft}, M., {Vazza}, F., {Br{\"u}ggen}, M., \&
  {Dom{\'\i}nguez-Fern{\'a}ndez}, P. 2019, \mnras, 490, 3987

\bibitem[{{Wittor} {et~al.}(2017){Wittor}, {Vazza}, \&
  {Br{\"u}ggen}}]{2017wittor}
{Wittor}, D., {Vazza}, F., \& {Br{\"u}ggen}, M. 2017, Monthly Notices of the
  Royal Astronomical Society, 464, 4448

\bibitem[{{Zwicky} {et~al.}(1961){Zwicky}, {Herzog}, {Wild}, {Karpowicz}, \&
  {Kowal}}]{1961zwicky}
{Zwicky}, F., {Herzog}, E., {Wild}, P., {Karpowicz}, M., \& {Kowal}, C.~T.
  1961, {Catalogue of galaxies and of clusters of galaxies, Vol. I}

\end{thebibliography}

\begin{appendix} %First appendix
\onecolumn
\section{Flux density for the beam depolarization}\label{app1}
\noindent
\begin{table*}[h!]
\small
\caption{Integrated and RM synthesis polarized flux densities and relative errors measured for the selected regions: north (\textit{top}) and south (\textit{bottom}).}
\label{tab:fluxes}
\centering
\begin{tabular*}{\textwidth}{c @{\extracolsep{\fill}} ccccccc}
\hline\hline
Beam size & $S_{\rm{P,int}}$ & $\delta S_{\rm{P,int}}$ & $S_{\rm{P,rm}}$ & $\delta S_{\rm{P,rm}}$ & $S_{\rm{I,int}}$ & $\delta S_{\rm{I,int}}$ \\
 & ($\rm{Jy}$) & ($\rm{Jy}$) & ($\rm{Jy}$) & ($\rm{Jy}$) & ($\rm{Jy}$) & ($\rm{Jy}$) \\
\hline
$13"~(57.9)$ & $7.40 \times 10^{-4}$ & $4.3 \times 10^{-5}$ & $1.135 \times 10^{-3}$ & $5.6 \times 10^{-5}$ & $6.109 \times 10^{-3}$ & $3.23 \times 10^{-4}$\\
\hline
$16"~(71.3)$ & $6.96 \times 10^{-4}$ & $5.7 \times 10^{-5}$ & $8.98 \times 10^{-4}$ & $6.8 \times 10^{-5}$ & $5.831 \times 10^{-3}$ & $4.45 \times 10^{-4}$\\
\hline
$20"~(89.0)$ & $6.40 \times 10^{-4}$ & $7.7 \times 10^{-5}$ & $7.31 \times 10^{-4}$ & $8.4 \times 10^{-5}$ & $5.466 \times 10^{-3}$ & $6.22 \times 10^{-4}$\\
\hline
$25"~(111.3)$ & $5.73 \times 10^{-4}$ & $1.14 \times 10^{-4}$ & $5.42 \times 10^{-4}$ & $1.01 \times 10^{-4}$ & $5.028 \times 10^{-3}$ & $8.61 \times 10^{-4}$\\
\hline
$30"~(133.5)$ & $5.23 \times 10^{-4}$ & $1.33 \times 10^{-4}$ & $5.31 \times 10^{-4}$ & $1.32 \times 10^{-4}$ & $4.612 \times 10^{-3}$ & $1.112 \times 10^{-3}$\\
\hline
$35"~(155.8)$ & $4.71 \times 10^{-4}$ & $1.60 \times 10^{-4}$ & $4.71 \times 10^{-4}$ & $1.57 \times 10^{-4}$ & $4.226 \times 10^{-3}$ & $1.366 \times 10^{-3}$\\
\hline
$40"~(178.0)$ & $4.28 \times 10^{-4}$ & $1.88 \times 10^{-4}$ & $4.25 \times 10^{-4}$ & $1.84 \times 10^{-4}$ & $3.847 \times 10^{-3}$ & $1.607 \times 10^{-3}$\\
\hline
$45"~(200.3)$ & $3.95 \times 10^{-4}$ & $2.17 \times 10^{-4}$ & $3.98 \times 10^{-4}$ & $2.11 \times 10^{-4}$ & $3.523 \times 10^{-3}$ & $1.849 \times 10^{-3}$\\
\hline
$50"~(222.5)$ & $3.63 \times 10^{-4}$ & $2.45 \times 10^{-4}$ & $3.48 \times 10^{-4}$ & $2.32 \times 10^{-4}$ & $3.252 \times 10^{-3}$ & $2.095 \times 10^{-3}$\\
\hline
\end{tabular*}
\quad
\begin{tabular*}{\textwidth}{c @{\extracolsep{\fill}} ccccccc}
\hline
Beam size & $S_{\rm{P,int}}$ & $\delta S_{\rm{P,int}}$ & $S_{\rm{P,rm}}$ & $\delta S_{\rm{P,rm}}$ & $S_{\rm{I,int}}$ & $\delta S_{\rm{I,int}}$ \\
 & ($\rm{Jy}$) & ($\rm{Jy}$) & ($\rm{Jy}$) & ($\rm{Jy}$) & ($\rm{Jy}$) & ($\rm{Jy}$) \\
\hline
$13"~(57.9)$ & $6.13 \times 10^{-4}$ & $2.9 \times 10^{-5}$ & $1.550 \times 10^{-3}$ & $4.8 \times 10^{-5}$ & $1.0598 \times 10^{-2}$ & $3.59 \times 10^{-4}$\\
\hline
$16"~(71.3)$ & $5.44 \times 10^{-4}$ & $3.5 \times 10^{-5}$ & $1.119 \times 10^{-3}$ & $5.1 \times 10^{-5}$ & $1.0404 \times 10^{-2}$ & $5.06 \times 10^{-4}$\\
\hline
$20"~(89.0)$ & $4.96 \times 10^{-4}$ & $4.4 \times 10^{-5}$ & $8.28 \times 10^{-4}$ & $5.9 \times 10^{-5}$ & $1.0142 \times 10^{-2}$ & $7.33 \times 10^{-4}$\\
\hline
$25"~(111.3)$ & $4.13 \times 10^{-4}$ & $5.0 \times 10^{-5}$ & $5.23 \times 10^{-4}$ & $5.6 \times 10^{-5}$ & $9.805 \times 10^{-3}$ & $1.064 \times 10^{-3}$\\
\hline
$30"~(133.5)$ & $4.12 \times 10^{-4}$ & $6.9 \times 10^{-5}$ & $5.42 \times 10^{-4}$ & $8.4 \times 10^{-5}$ & $9.446 \times 10^{-3}$ & $1.439 \times 10^{-3}$\\
\hline
$35"~(155.8)$ & $3.80 \times 10^{-4}$ & $8.2 \times 10^{-5}$ & $4.86 \times 10^{-4}$ & $1.00 \times 10^{-4}$ & $9.072 \times 10^{-3}$ & $1.850 \times 10^{-3}$\\
\hline
$40"~(178.0)$ & $3.51 \times 10^{-4}$ & $9.6 \times 10^{-5}$ & $4.42 \times 10^{-4}$ & $1.17 \times 10^{-4}$ & $8.660 \times 10^{-3}$ & $2.278 \times 10^{-3}$\\
\hline
$45"~(200.3)$ & $3.21 \times 10^{-4}$ & $1.07 \times 10^{-4}$ & $2.02 \times 10^{-4}$ & $1.33 \times 10^{-4}$ & $8.279 \times 10^{-3}$ & $2.730 \times 10^{-3}$\\
\hline
$50"~(222.5)$ & $2.97 \times 10^{-4}$ & $1.20 \times 10^{-4}$ & $3.97 \times 10^{-4}$ & $1.59 \times 10^{-4}$ & $7.909 \times 10^{-3}$ & $3.190 \times 10^{-3}$\\
\hline
\label{tab:smooth}
\end{tabular*}
\tablefoot{Column 1: beam size for the different Gaussian kernel (with the equivalent scale in kpc in brackets). Column 2: polarized flux density in integrated analysis $S_{\rm{P,int}}$. Column 3: $\delta S_{\rm{P,int}}$ obtained as seen in Eq.~\ref{eq:fluxerror}. Column 4: polarized flux density with RM synthesis $S_{\rm{P,rm}}$. Column 5: error on flux density $\delta S_{\rm{P,rm}}$. Column 6: total intensity flux density $S_{\rm{I, int}}$. Column 7: error on flux density $\delta S_{\rm{I, int}}$}
\end{table*}
\section{Analysis and discussion on the nature of complex Faraday spectra}\label{app2}
With the RM synthesis it is possible to obtain the reconstructed (and eventually cleaned) FDF, also called Faraday spectrum, that shows for each pixel the trend of polarized flux as a function of the Faraday depth. In the simplest case in which we observe a source with an intrinsic rotation measure whose radiation crosses a non-emitting magnetized plasma, we expect to observe a $\tilde{F}(\phi)$ with a single peak, corresponding to the resulting RM value. A good example of this case is represented in Fig.~\ref{fig:rmsf_and_fdf}. However, there can be some pixels in which the reconstructed FDF shows multiple peaks: this can happen, for example, when several contributions mixes along the same line of sight (e.g., a linearly polarized source whose radiation crosses an emitting magnetized plasma, or the alignment of multiple polarized emitting sources), and in these cases the correct reconstruction of the true RM value can be challenging. We usually point out to these as complex Faraday spectra.
\par
\begin{figure}[h!]
\centering
\includegraphics[width=\columnwidth]{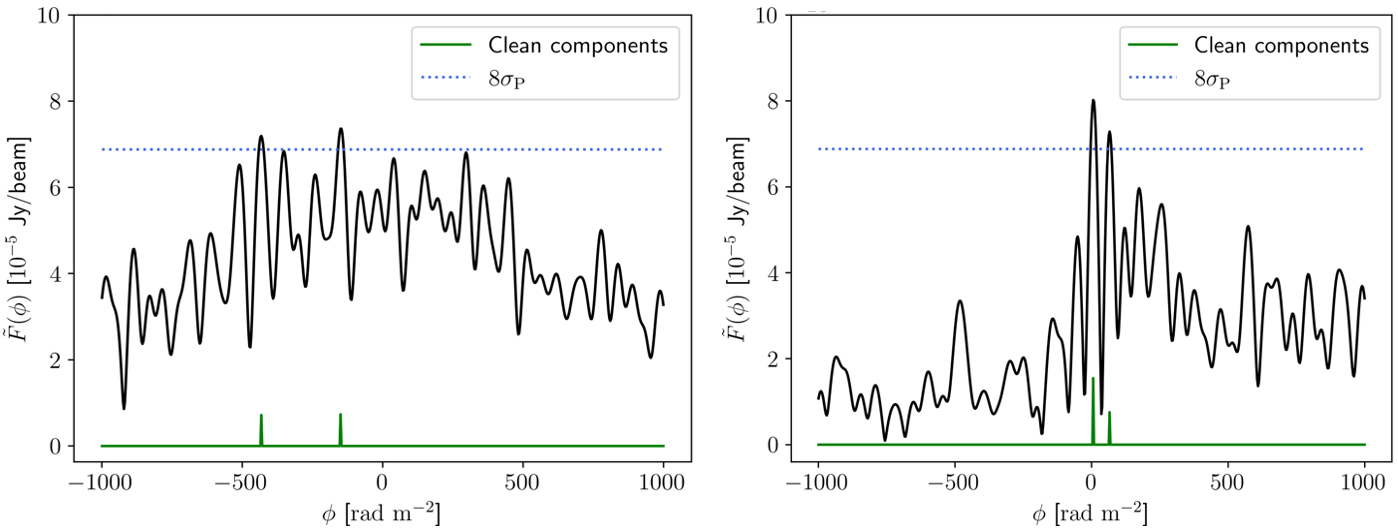}
\caption{Example of complex Faraday spectrum for three pixels. The black solid line represents the trend of the cleaned, reconstructed FDF as a function of the Faraday depth $\phi$. The horizontal blue dotted line represents the threshold used to select the clean components for the \texttt{RMclean3D} of $8\sigma_{\rm{P}}$ (in blue), with $\sigma_{\rm{P}} = 8~\rm{\mu Jy~beam^{-1}}$.}
\label{fig:complex_fdf}
\end{figure}

 Through our RM synthesis analysis of the relics, we found several pixels (about 6\%) showing a complex Faraday spectrum, especially for the north relic. Some examples are shown in Fig.~\ref{fig:complex_fdf}. Usually the presence of such behavior in FDF could mean that there are several contribution to the observed RM along the same line of sight, and so the assumption that we only have a non-emitting Faraday screen would not be valid. However, analyzing all the complex pixels, we notice that they are mainly located at the boundary of the regions selected for the relics and, moreover, they are predominantly dominated by noise. Because of these evidences, we can conclude that they are not to be taken into account for further analysis about their complex topology, that would imply additional features for the reconstruction of the correct RM value.

\end{appendix}
\end{document}